\documentclass{ws-ijmpd}
\usepackage[super,compress]{cite}
\usepackage[breaklinks]{hyperref}
\hypersetup{colorlinks,urlcolor=black,citecolor=black,linkcolor=black,filecolor=black}
\usepackage{breakurl}

\begin{document}

\markboth{C. Burigana et al.}
{Cosmic backgrounds from the radio to the far-infrared}


%
\catchline{}{}{}{}{}
%


\title{Cosmic backgrounds from the radio to the far-infrared: recent results \\ and perspectives from cosmological and astrophysical surveys
            }
            
\author{Carlo Burigana;$^{1,2,3,a}$ 
Elia Sefano Battistelli;$^{4,5,b}$
Laura Bonavera;$^{6,7,c}$\\
Tirthankar Roy Choudhury;$^{8,d}$
Marcos Lopez-Caniego;$^{9,e}$
Constantinos Skordis;$^{10,f}$\\
Raelyn Marguerite Sullivan;$^{11,g}$
Hideki Tanimura;$^{12,h}$
Seddigheh Tizchang;$^{13,i}$\\
Matthieu Tristram;$^{14,j}$
Amanda Weltman$^{15,k}$
}
\address{$^{1}$INAF--IRA, Via Piero Gobetti 101, 40129 Bologna, Italy
\footnote{Istituto Nazionale di Astrofisica --
Istituto di Radioastronomia, Via Piero Gobetti 101, 40129 Bologna, Italy}\\
$^{2}$Dipartimento di Fisica e Scienze della Terra, Universit\`a degli Studi di Ferrara,\\Via Giuseppe Saragat 1, I-44122 Ferrara, Italy\\
$^{3}$INFN, Sezione di Bologna, Via Irnerio 46, 40126, Bologna, Italy\\
$^{4}$Dipartimento di Fisica, Universit\`a di Roma ``La Sapienza'',\\P.le Aldo Moro 2, 00185, Rome, Italy\\
$^{5}$INAF-IAPS Roma, Via del Fosso del Cavaliere 100, 00133 Roma, Italy\\
$^{6}$Departamento de F\'{i}sica, Universidad de Oviedo,\\C. Federico Garc\'{i}a Lorca 18, 33007 Oviedo, Spain\\
$^{7}$Instituto Universitario de Ciencias y Tecnolog\'ias Espaciales de Asturias (ICTEA),\\C. Independencia 13, 33004 Oviedo, Spain\\
$^{8}$National Centre for Radio Astrophysics, Tata Institute of Fundamental Research, Ganeshkhind, Pune 411007, India\\
$^{9}$Aurora Technology B.V. for the European Space Agency,\\Villanueva de la Canada, Madrid, 28692, Spain \\
$^{10}$CEICO, FZU - Institute of Physics of the Czech Academy of Sciences,\\ Slovance 1999/2, 182 21, Prague, Czech Republic \\
$^{11}$Department of Physics \& Astronomy, University of British Columbia,\\6224 Agricultural Road, Vancouver, British Columbia, Canada\\
$^{12}$Universit\'{e} Paris-Saclay, CNRS, Institut d'Astrophysique Spatiale,\\B\^atiment 121, 91405 Orsay, France \\
$^{13}$School of Particles and Accelerators, Institute for Research in Fundamental Sciences (IPM),\\PO BOX19395-5531, Tehran, Iran\\
$^{14}$Universit\'e Paris-Saclay, CNRS/IN2P3, IJCLab, 91405 Orsay, France\\
$^{15}$High Energy Physics, Cosmology \& Astrophysics Theory Group, University of Cape Town,\\Private Bag, Cape Town, South Africa, 7700
\vskip 0.1cm
$^{a}$burigana@ira.inaf.it $-$ $^{b}$elia.battistelli@roma1.infn.it $-$ $^{c}$bonaveralaura@uniovi.es\\
$^{d}$tirth@ncra.tifr.res.in $-$ $^{e}$marcos.lopez.caniego@sciops.esa.int $-$ $^{f}$skordis@fzu.cz\\
$^{g}$rsullivan@phas.ubc.ca $-$ $^{h}$hideki.tanimura@ias.u-psud.fr $-$ $^{i}$s.tizchang@ipm.ir\\
$^{j}$tristram@ijclab.in2p3.fr $-$ $^{k}$amanda.weltman@uct.ac.za
}

\maketitle

\begin{history}
\received{24 November 2021}
\revised{23 December 2021}
\accepted{1 January 2022}
\end{history}

\begin{abstract}
Cosmological and astrophysical 
surveys in various wavebands, in particular
from the radio to the far-infrared, 
offer a unique view of the universe's properties and the formation and evolution of its structures.
After a preamble on the so-called tension problem, which occurs when different types of data are used to determine cosmological parameters, we discuss the role of fast radio bursts in cosmology,
in particular for the missing baryon problem,
and the perspectives from the analysis of the 21~cm redshifted line from neutral hydrogen.
We then describe the {\it Planck} Legacy Archive,
its wealth of scientific information and next developments, and the promising perspectives expected from higher resolution observations,
in particular for the analysis of the thermal Sunyaev-Zel'dovich effect.
Three cosmological results of the {\it Planck} mission are presented next: the implications of the map of Comptonization fluctuations, 
the dipole analysis from cross-correlating cosmic microwave background anisotropy and Comptonization fluctuation maps,
and the constraints on the primordial tensor-to-scalar perturbation ratio.
Finally, we discuss some future perspectives and alternative scenarios in cosmology, such as
the study of the Lorentz invariance violation 
with the cosmic microwave background polarization, the introduction of new gravitational degrees of freedom to solve the dark matter problem,
and the exploitation of the magnification bias with high-redshift sub-millimeter galaxies to constrain cosmological parameters.
\end{abstract}


\ccode{PACS numbers: 98.80.-k; 98.70.Vc; 98.80.Es; 95.85.Bh; 95.85.Fm; 98.38.Jw; 98.65.Dx; 98.65.Cw; 98.70.Dk; 98.70.Lt; 98.62.Sb; 95.30.Sf; 04.50.Kd; 95.35.+d; 03.30.+p; 95.30.Gv.}

\def\beq{\begin{equation}}
\def\eeq{\end{equation}}

\def\simlt{\mathrel{\rlap{\lower 3pt\hbox{$\sim$}}\raise 2.0pt\hbox{$<$}}}
\def\simgt{\mathrel{\rlap{\lower 3pt\hbox{$\sim$}} \raise
2.0pt\hbox{$>$}}}
\def\lsim{\,\lower2truept\hbox{${<\atop\hbox{\raise4truept\hbox{$\sim$}}}$}\,}
\def\gsim{\,\lower2truept\hbox{${> \atop\hbox{\raise4truept\hbox{$\sim$}}}$}\,}
\def\aap{A\&A}
\def\apj{ApJ}
\def\apjs{ApJS}
\def\apjl{ApJL}
\def\mnras{MNRAS}
\def\aj{AJ}
\def\nat{Nature}
\def\aaps{A\&A Supp.}
\def\aaps{A\&A Supp.}
\def\pra{Phys.Rev.A}         
\def\physrep{Physics Reports}         
\def\prb{Phys. Rev. B}         
\def\prc{Phys. Rev. C}         
\def\prd{Phys. Rev. D}         
\def\prl{Phys. Rev. Lett.}      
\def\araa{ARA\&A}       
\def\gca{GeCoA}         
\def\pasp{PASP}              
\def\pasj{PASJ}              
\def\apss{ApSS}
\def\jcap{JCAP}
\def\sovast{Soviet Astronomy}
\def\na{New Astronomy}
\def\aapr{A\&A Rev.}
\def\planss{Planet. Space Sci.}
\def\procspie{Proceedings of the SPIE}

\def\Planck{\textit{Planck}}

\providecommand{\sorthelp}[1]{}
\let\vec\mathbf
\def\WMAP{\textit{WMAP}}
\def\COBE{\textit{COBE}}
\def\LCDM{$\Lambda$CDM}
\def\smica{{\tt SMICA}}
\def\smicano{{\tt SMICA-NOSZ}}
\def\ilc{{\tt ILC}}
\def\milca{{\tt MILCA}}
\def\twodilc{{\tt 2D-ILC}}
\def\Firas{\textit{Firas}}
\def\aap{A\&A}%
\newcommand{\dk}[1]{\textcolor{red}{#1}}
\newcommand{\dktwo}[1]{\textcolor{blue}{#1}}

\def\gnilc{{\tt GNILC}}
\def\sevem{{\tt SEVEM}}
\def\commander{{\tt Commander}}
\def\planckskymodel{{\tt Planck Sky Model}}
\def\healpix{{\tt HEALPix}}


\def\planck{\textit{Planck}}
\def\lcdm{$\rm{\Lambda CDM}$}
\providecommand{\Omk}{\Omega_K}
\providecommand{\Oml}{\Omega_{\Lambda}}
\providecommand{\Omtot}{\Omega_{\mathrm{tot}}}
\providecommand{\Omb}{\Omega_{\mathrm{b}}}
\providecommand{\Omc}{\Omega_{\mathrm{c}}}
\providecommand{\Omm}{\Omega_{\mathrm{m}}}
\providecommand{\omb}{\omega_{\mathrm{b}}}
\providecommand{\omc}{\omega_{\mathrm{c}}}
\providecommand{\omm}{\omega_{\mathrm{m}}}
\providecommand{\Omdm}{\Omega_{\mathrm{DM}}}
\providecommand{\Omnu}{\Omega_{\nu}}
\newcommand{\ns}{n_{\rm s}}
\newcommand{\Mpc}{\text{Mpc}}
\def\kmsMpc{\ifmmode $\,\kms\,Mpc\mo$\else \,\kms\,Mpc\mo\fi}
\newcommand{\Hunit}{~\text{km}~\text{s}^{-1} \Mpc^{-1}}
\newcommand{\nnu}{N_{\rm eff}}
\newcommand{\mnu}{\sum m_\nu}
\newcommand{\eV}{\,\text{eV}}
\newcommand{\wzero}{w_0}
\newcommand{\yhe}{Y_{\text{P}}}
\newcommand{\planckonly}{\planck}
\newcommand{\zre}{z_{\text{re}}}
\newcommand{\thetastar}{\theta_{\ast}}
\newcommand{\rstar}{r_{\ast}}
\newcommand{\DM}{D_{\rm M}}

\section{Introduction}\label{sect:intro}

Over the last century or so, cosmology has progressed from a field that was essentially philosophy to a precision science.
The current standard cosmological model is the $\Lambda$ cold dark matter ($\Lambda$CDM) model, which
mainly consists of a cosmological constant ($\Lambda$) or dark energy (DE) component and cold dark matter (CDM) component,
with general relativity (GR) as the assumed theory of gravity. It makes predictions on the smallest and largest scales, that, 
despite some open questions, are largely verified by a host of observational cosmology experiments. 

Cosmological and astrophysical 
surveys in various wavebands, in particular
from the radio to the far-infrared, 
are crucial to solve many problems in cosmology
and to understand the origin and evolution of the cosmic structures at various
scales through cosmic time. 
The accurate observations of the properties of the cosmic microwave background (CMB) 
began the era of precision cosmology,
furthering our understanding of the early universe and its evolution up to the present time and beyond. 
Similarly, the information contained in the radio sky is of increasing relevance to unravel the complexity 
of cosmic evolution and to answer a wide range of open questions, 
thanks to the spectacular experimental improvement achieved in the past years and expected in the future.

The structure of this paper is as follows.
In Sect. \ref{sect:cosmo_radio}, we begin with a preamble to the so-called tension problem, which occurs when different data are used to determine cosmological parameters. We then discuss the role of fast radio bursts (FRBs) in cosmology to solve the missing baryons problem, followed by the perspectives represented by the analysis of the 21~cm redshifted line from neutral hydrogen (HI). 
Sect. \ref{sect:obs_microwaves} describes the wealth of scientific information and tools publicly available at the \Planck\ Legacy Archive (PLA), with its current and future developments, and the perspectives expected from higher resolution observations at millimeter (mm) and sub-mm wavelengths, in particular for the analysis of the thermal Sunyaev-Zel'dovich (SZ) effect towards galaxy clusters (GCs), focusing on a forthcoming well-defined facility instrument.
In Sect. \ref{sect:PlanckCosmo} we discuss three important cosmological results based on the data from the \Planck\ mission: the implications of the analysis of the thermal SZ (tSZ) map, 
the dipole analysis based on the cross-correlation of CMB anisotropy and tSZ maps, and the constraints on the ratio between
the primordial tensor and scalar perturbations.
Finally, Sect. \ref{sect:future_alternat} is dedicated to future perspectives and alternative scenarios in cosmology. We consider three specific examples: the study of Lorentz invariance violation 
with the analysis of CMB polarization, the solution to the dark matter (DM) problem introducing new gravitational degrees of freedom, and the exploitation of the magnification bias with
high-$z$ sub-mm galaxies to constrain a set of cosmological parameters.

\section{Cosmology in the radio}\label{sect:cosmo_radio}

\subsection{Cosmology with radio astrophysics and tensions in cosmology}\label{subsect:cosmo_radioastroph}

Amongst the successes of modern cosmology, is the remarkable precision in the estimation of the cosmological parameters that describe the contents, geometry and history of our universe.
Herein also lies the current greatest tension in the data. 
Indeed, in spite of the increasing precision in observational cosmology, 
it is possible we have lost some accuracy:
a number of tensions now appear within our data, tensions that more cosmology experiments do not appear to solve. These tensions may require a 
change in the underlying theory 
or they may be resolved with novel observational probes of the large scale picture of the universe.
Here we will describe an example of the latter, namely, the FRBs, 
the novel discovery of our century, as potential instruments for cosmology. 
Indeed we will highlight more broadly the use of radio astrophysics - encompassing both the use of radio transients and intensity mapping tools, and the potential they hold for groundbreaking cosmology discovery.

The Hubble constant, $H_0$, is anything but constant across observations at different scales
(see Ref.~\citenum{DiValentino:2021izs} for a recent review of most of the data).
A brief summary however, is that historically there were two independent measures for $H_0$,
very far apart in value with measurements of roughly 50\,km\,s$^{-1}$\,Mpc$^{-1}$ or 100\,km\,s$^{-1}$\,Mpc$^{-1}$ from the early and late universe respectively. However, the error bars on each measurement was large enough, that each was within errors of the other and it was assumed that with improved observations, the error bars would get smaller and the values would get closer and converge. Indeed, this appeared to be the case early this century, but every new data set over the last several years has the values diverging with ever smaller error bars, that critically no longer allow for a simple convergence in the value of $H_0$.
Measurements from the early universe,
such as \Planck, 
Dark Energy Survey\cite{DES2018,DES2021} (DES),
and baryonic acoustic oscillations (BAO) 
and Big Bang nucleosynthesis (BBN) 
observations,
all point to a smaller
value
of around $H_0 \simeq 67$\,km\,s$^{-1}$\,Mpc$^{-1}$
and measurements from the late universe, using Cepheids as distance ladders, or lensing for example, find
$H_0 \simeq 73$\,km\,s$^{-1}$\,Mpc$^{-1}$
with each measurement having error bars
in the range $\simeq$1$\div$4\,km\,s$^{-1}$\,Mpc$^{-1}$.
Thus this tension is growing with data rather than resolving.

There are longer standing issues with the standard cosmological model, of course, the lack of direct observational evidence for DM which makes up roughly 25\% of the total energy budget of the universe is of particular concern. No less resolved is the issue of DE, here making up $\sim70\%$ of the energy budget of the universe and yet only observed in indirect cosmological experiments with no direct detection and also no compelling theoretical explanation for why it even exists or why it appears to be dominating the total energy budget of the universe at a cosmological time when we are both possible and here to observe it.
{Early universe inflation answered several questions, such as the flatness, horizon and primordial monopole
problems,\cite{1978AnPhy.115...78B,1980PhLB...91...99S,1980ApJ...241L..59K,1981PhRvD..23..347G,1981MNRAS.195..467S,1982PhLB..108..389L,1982PhRvL..48.1220A,1983PhLB..129..177L}
and is one of the viable scenarios predicting perturbations consistent with a nearly scale-invariant power spectrum,\cite{SZ_1970a,SZ_1970b,Harrison_1970,PY_1970,Z_1972} which is related, in this framework,
to quantum fluctuations (see Sect. \ref{subsect:Planck_r} for references),
and yet we still have no theory of inflation that fits within an ultraviolet complete theory with a natural candidate for the inflaton lacking.}
One hopes that these three problems are resolved together with a leap in the theory space - yet the obligation remains to continue to search for observational data that may yet play a role.

Other problems have lingered for very long, such as the missing baryon problem, where even the $\sim5\%$ of the universe that we know exists
is not entirely accounted for.
Indeed adding up all of the observed contributions to the baryons accounts for only $\sim70\%$ of the total expected. Roughly 30\% are considered missing, and expected to be somewhere in the warm hot intergalactic medium (WHIM).
Indeed, we should say were considered missing, as we discuss below how to find these missing baryons using 
the FRBs. 

\subsubsection{Fast radio bursts and their role in cosmology}

As the name suggests, FRBs are very bright ($\sim$\,Jy), brief ($\sim$$\mu$s to $\sim$\,ms scale) transients, observed in the broad spectrum of the radio. The first was discovered\cite{Lorimer:2007qn} in 2007 and ever since the search has been on to discover more and their properties, first through searches in archival data and more recently through purpose built radio telescopes and arrays to find thousands if not tens of thousands of bursts every year. Our understanding of the properties of FRBs is still evolving; it appears that many but not all repeat, it is not yet known if any of the repeaters repeat in a periodic fashion, and indeed it is likely that ultimately FRBs will fall in several classes as has happened with transients historically. Their other properties such as polarisation, rotation measures etc are all still unknown as the observations do not all point to a single pattern and it is not always possible to disentangle host galaxy effects from propagation effects from intrinsic properties of the bursts. It does appear certain that they are found in host galaxies, and that they lie at cosmological distances,\cite{Tendulkar:2017vuq} thus making them excellent candidates for cosmology as their propagation will probe the intergalactic medium (IGM) and thus give us a number of novel ways to constrain cosmological parameters (see, e.g., Refs.~\citenum{Walters:2017afr,Walters:2019cie,Weltman:2019cqv} and references therein). From a theoretical standpoint, the progenitor mechanism driving FRBs is not yet known though there are strong hints that at least some are driven by magnetars through one of many physical possible mechanisms. A full database of theories, \cite{Platts:2018hiy} once outnumbered the observations, but as the amount of data grows, the possibilities for theoretical explanations shrinks and so it is likely that in a few years we will have only a handful of possible contenders remaining, perhaps matching future classes of FRBs. 

There are a number of ways FRBs can be used as cosmological probes. Here we focus on a single application. After emission from the source, the photons from the burst travel to the observer through the IGM, and are slowed down differentially as a function of their wavelength, $\lambda$, or frequency, $\nu=c/\lambda$, $c$ being the speed of light.
The dispersion measure, $M_D \simeq \int n_e dl$, 
where $n_e$ is the electron number density and the integral along the line of sight 
runs from the source to the observer,
thus contains cosmological information about the distribution of electrons
Indeed, this simple equation combined with precision cosmology constraints from the early universe allows us to make a prediction for the fraction of baryons in the IGM,\cite{Walters:2019cie} which can then be experimentally verified. Indeed, the so-called missing baryon problem is not longer an outstanding problem as the results of Ref.~\citenum{Macquart:2020lln} used a handful of well located FRBs with the observed dispersion measures to show that indeed the missing baryons are in the IGM as expected,
and are 
playing 
the predicted
role of dispersing the FRB signal. This longstanding open problem is thus resolved not due to any great technological or theoretical breakthrough, but simply through the use of a few transient observations and an understanding of the contributions to $M_D$
from our own galaxy.
This suggests that there is immense and untapped potential for great discovery with FRBs and that the future is very bright for this young field.

\subsection{High-redshift universe with redshifted 21~cm line}\label{subsect:highz_univ_21cm}

The redshifted 21~cm line of HI is one of the most useful probes of the early universe. Several experiments are ongoing and are being planned to detect the signal 
at high redshift, $z$.
Detection of the signal will help in understanding the first stars in the universe, the formation and evolution of galaxies and also constraining cosmological parameters.
This section summarizes the status of observational efforts and theoretical understanding.

The HI evolution can be used to study the thermal and ionization history of the universe and also the evolution of the galaxies.\cite{2001PhR...349..125B,2009CSci...97..841C}
The very first HI atoms formed during the so-called `recombination' epoch.
The hydrogen remains in the neutral form until the first stars form which can then start ionizing the HI in the IGM, a process known as \emph{reionization}. This epoch is believed to end around redshifts $z \sim 5.5$ by when almost all the HI in the IGM gets ionized. In the \emph{post-reionization} era, the only regions where HI can survive are the high-density regions, e.g., the interstellar medium (ISM) of the galaxies.

The evolution of HI can be efficiently tracked by the redshifted 21~cm signal arising from hyperfine transition of the ground state of the atom. The signal is observed, either in emission or absorption, in contrast to the CMB signal using low-frequency radio telescopes. Since the hyperfine transition is forbidden, the signal amplitude turns out to be extremely weak and hence is appreciable only when the HI density is significant. There are broadly three regimes where the detection of the 21~cm radiation is being planned:
(i) the \emph{cosmic dawn} when the ultraviolet (UV) and X-ray radiation from the very early galaxies lead to a strong absorption signal around $z \sim 20$;
(ii) the \emph{epoch of reionization} when the heated and ionized IGM leads to an emission signal around $z \sim 15-6$; and (iii) the \emph{post-reionization} signal when HI in galaxies can be detected either directly or through intensity mapping.

\begin{figure}[b]
         \vskip -0.4cm
\begin{center}
\includegraphics[width=0.52\textwidth]{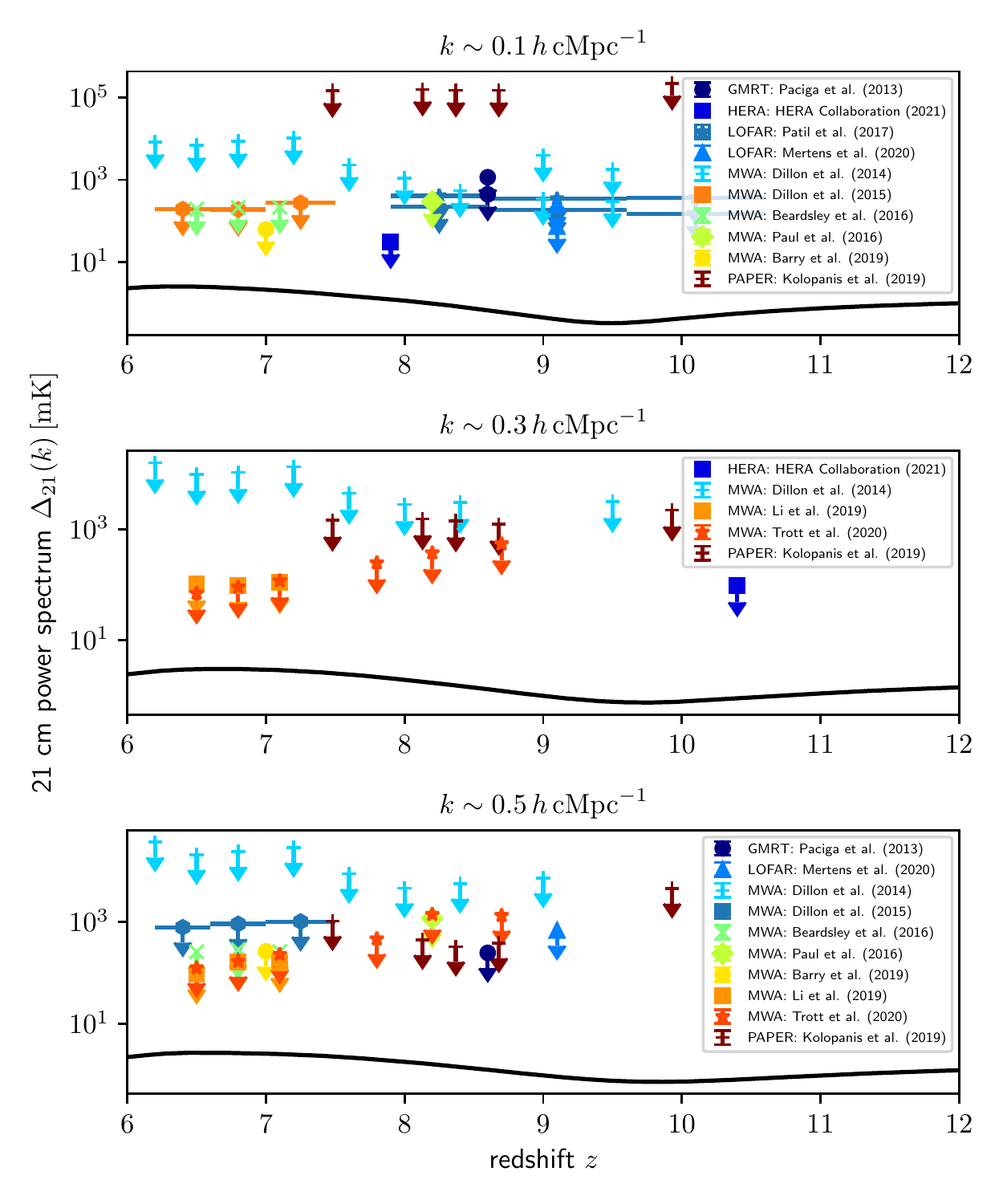}
\end{center}
      \vskip -0.5cm
\caption{Observational upper limits 
to
the 21~cm PS, $\sqrt{k^3 P_{21}(k) / 2  \pi^2}$, for three different comoving wavenumbers
$k$; here $h = H_0 /$\,(100\,km\,s$^{-1}$\,Mpc$^{-1}$) and cMpc denotes comoving Mpc.
Data from different telescopes, 
namely,
the Giant Metrewave Radio Telescope\cite{2013MNRAS.433..639P} (GMRT),
the Low Frequency Array\cite{2017ApJ...838...65P,2020MNRAS.493.1662M} (LOFAR),
the Murchison Widefield Array\cite{2014PhRvD..89b3002D,2015PhRvD..91l3011D,2016ApJ...833..102B,2016ApJ...833..213P,2019ApJ...884....1B,2019ApJ...887..141L,2020MNRAS.493.4711T} (MWA),
the Precision Array for Probing the Epoch of Reionization\cite{2019ApJ...883..133K} (PAPER), 
the Hydrogen Epoch of Reionization\cite{2021arXiv210802263T} Array (HERA),
are shown
in comparison with
the signal predictions from
the Semi-numerical Code for ReIonization with PhoTon-conservation\cite{2018MNRAS.481.3821C} (\texttt{SCRIPT}).}
\label{fig:21cmPk}
         \vskip -0.4cm
\end{figure}

Among the different ways to detect the 21~cm signal, perhaps the one that requires simplest of the instruments is the signal averaged over large regions of the sky, i.e., the \emph{global signal}. This signal is believed to be most prominent during the cosmic dawn corresponding to 
$\nu \simeq 50-100$~MHz.
The amplitude of the cosmological signal is weak and is buried under other astrophysical signals few orders of magnitude stronger, which makes it extremely challenging to separate it out.

Recently, there has been a claim of a detection of the 21~cm signal from cosmic dawn 
at $z \sim 18$
by the 
Experiment to Detect the Global EoR Signature\cite{2018Natur.555...67B} (EDGES),
although the recovered signal seems to be $3-4$ times larger than that predicted by standard galaxy formation models. This lead to several theoretical interpretation beyond the standard calculations, e.g., exotic physics in the DM sector\cite{2018Natur.555...71B} or strong radio background from the first galaxies.\cite{2018ApJ...858L..17F,2019MNRAS.486.1763F,2020MNRAS.496.1445C} There have also been concerns regarding the foreground subtraction and other systematics in the data,\cite{2018Natur.564E..32H} hence it is important to validate the signal through other low-frequency telescopes.

An alternate way of detecting the 21~cm signal is through the spatial fluctuations. These can, in principle, be useful throughout the cosmic history starting from the cosmic dawn to the post-reionization universe. During the cosmic dawn, the fluctuations arise from those in the galactic radiation, while in the epoch reionization it is the patchiness in the ionization field that drives the 21~cm fluctuation signal. In the
post-reionization era, the fluctuations in the 21~cm signal essentially trace the distribution of galaxies and hence the underlying matter fluctuations.

At present, a number of low-frequency interferometers are attempting to detect the 21~cm power spectrum (PS) from the epoch of reionization
at scales $\sim 10 h^{-1}$\,cMpc. 
Fig. \ref{fig:21cmPk} shows the current upper limits
(see Ref. \citenum{2021MNRAS.507.4684R} for a complete compilation of available data).
These observational attempts are ably complemented by numerical and semi-numerical theoretical models to help interpret the data. The result from the one such semi-numerical simulation \texttt{SCRIPT}\footnote{\url{https://bitbucket.org/rctirthankar/script}} is shown in Fig. \ref{fig:21cmPk} for a model of reionization consistent with all other available observations.\cite{2018MNRAS.481.3821C}
Upcoming facilities like the Square Kilometre Array\footnote{\url{https://www.skatelescope.org}} (SKA) will be able to reach a noise level of $\sim 1$~mK ($\sim 0.3$~mK) within $\sim 100$ ($1000$) hours of observations and hence should constrain reionization models with high statistical significance.

\section{Observational results and perspectives in the microwaves}\label{sect:obs_microwaves}

\subsection{The \textit{Planck} Legacy Archive, present and future}\label{subsect:PLA}

\Planck\ was an ESA space
satellite that measured the microwave sky in nine wavebands, allowing it to not only capture the CMB to incredible precision, but also to 
accurately study 
a number of astrophysical diffuse emissions and discrete sources
(see fig.~51 in Ref.~\citenum{planck2014-a12}, which shows the various wavebands of the \Planck\ satellite, as well as the foregrounds signals and their frequency spectra across those bands). 

\begin{figure}[t!]
    \centering
    \includegraphics[width=1.\linewidth]{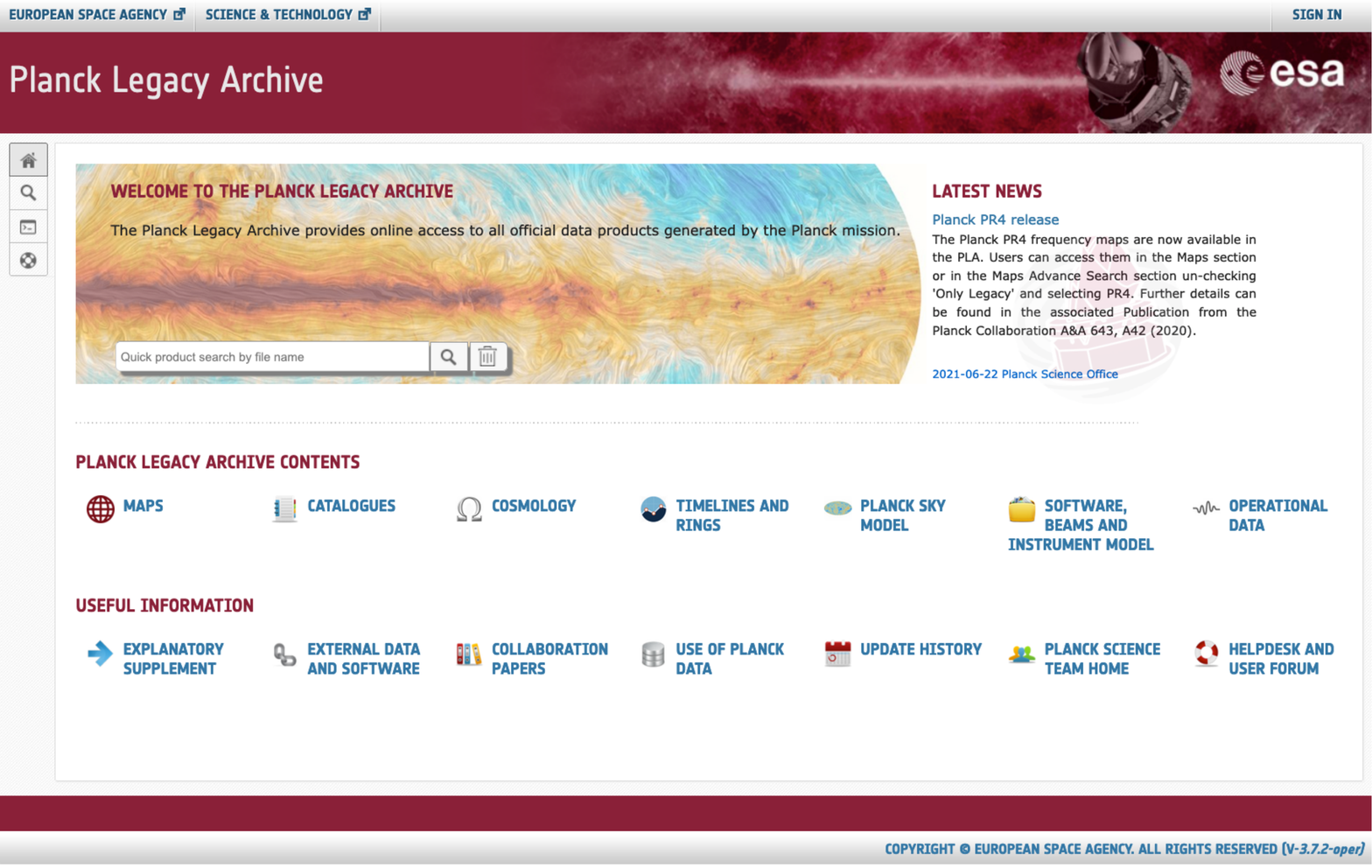}
        \vskip -0.4cm
        \caption{The \Planck\ Legacy Archive home page.}
    \label{fig:pla}
     \vskip -0.1cm
    \end{figure}

The PLA hosts and serves over 150 TB of products from \Planck\, that are publicly
available
via the PLA web interface\footnote{\url{https://pla.esac.esa.int}}
(Fig. \ref{fig:pla}).
This 
interface provides direct access to the users to a wide variety of products  
of the \Planck\ Collaboration\footnote{The \Planck\ Collaboration acknowledges the support of: ESA; CNES, and
CNRS/INSU-IN2P3-INP (France); ASI, CNR, and INAF (Italy); NASA and DoE
(USA); STFC and UKSA (UK); CSIC, MINECO, JA, and RES (Spain); Tekes, AoF,
and CSC (Finland); DLR and MPG (Germany); CSA (Canada); DTU Space
(Denmark); SER/SSO (Switzerland); RCN (Norway); SFI (Ireland);
FCT/MCTES (Portugal); ERC and PRACE (EU). A description of the \Planck\
Collaboration and a list of its members, indicating which technical
or scientific activities they have been involved in, can be found at
\url{http://www.cosmos.esa.int/web/planck/planck-collaboration}.}
through the Low Frequency Instrument (LFI) Data Processing Centre (DPC) in Trieste, Italy, by the High Frequency Instrument (HFI) DPC in Paris, France, and, more recently,
by the US \Planck\ Data Center in Pasadena and Berkeley, California, USA.
The PLA is located at the European Space Astronomy Centre (ESAC) near Madrid, Spain, and is maintained by the Data Science and Archives Division personnel.

In the last years, four major releases of \Planck\ products have taken place, providing users with the most complete cosmology data set to date.

The \Planck\ products can be divided into various categories.
{\it First}, the time-ordered data,\footnote{\url{https://pla.esac.esa.int/\#timelines}}
that comes in multiple flavors: raw, semi-calibrated and calibrated science ready timelines and compressed timelines (called HPRs).
{\it Second}, the frequency maps produced by the DPCs, in most cases for each detector, detector pair and frequency, and for different combinations of the mission coverage
(e.g., yearly maps, nominal mission, half-mission, half-ring, full mission, etc);
{\it third}, the maps of the CMB;
and, {\it fourth}, the so-called astrophysical component maps
that contain the diffuse emission from our Galaxy as produced by the dust grains, synchrotron emission from
spiraling electrons moving along the magnetic field lines, 
cosmic infrared background (CIB), etc. 
These {frequency}, CMB and astrophysical maps,
easily accessible through a set of dedicated web pages,\footnote{\url{https://pla.esac.esa.int/\#maps}} 
are derived in the scheme offered by the
Hierarchical Equal Area isoLatitude Pixelization of a sphere\cite{2005ApJ...622..759G} (\healpix) 
using different component separation methods, e.g., 
\commander, 
Spectral Matching Independent Component Analysis (\smica), 
Spectral Estimation Via Expectation Maximisation (\sevem) and  Generalized Needlet Internal Linear Combination (\gnilc),
the four official component separation methods used to extract the CMB maps in the third \Planck\ Release (PR3).
{\it Fifth}, source catalogues in various forms,\footnote{\url{https://pla.esac.esa.int/\#catalogues}} from Galactic and extragalactic compact source catalogues, available in the \Planck\ Catalogue of Compact Sources (PCCS), 
to galactic cold cores, SZ GCs, high-$z$ lensing candidates, etc.
{\it Sixth}, cosmology products,\footnote{\url{https://pla.esac.esa.int/\#cosmology}}  e.g., the \Planck\ likelihood code and associated files, the different flavors of the \Planck\ CMB angular
PS
as a function of the multipole\footnote{{The multipole $\ell$ is essentially the Fourier conjugate of the angular scale $\theta$: they are approximately related by
$\ell \simeq 180 / \theta$, with $\theta$ expressed here in degrees.}}
$\ell$
(e.g., low-$\ell$, high-$\ell$, temperature-only, temperature and polarization, etc), lensing products and more.
    
Advance Search panels are available to extensively query the PLA database, in addition to embedded links to the \Planck\ Explanatory Supplement
documentation,\footnote{\url{https://wiki.cosmos.esa.int/planck-legacy-archive}} multiple data download options, and Helpdesk support.

Three major releases of \Planck\ products took place in 2013, 2015, and 2018 and a selection of products have been tagged as
``Legacy''
to identify the version of each product most appropriate for general scientific use. In 2021 a new release of products will take place with a joint reprocessing of LFI+HFI time-ordered data that includes additional information not used in previous releases. In addition, EU funded projects reprocessing \Planck\ data, or combining it with other experiments, are expected to the deliver to the PLA higher level data products of interest.
The PLA also offers specialized tools\footnote{\url{https://pla.esac.esa.int/\#docsw}}
 that facilitate the processing of \Planck\ products. These tools are mainly designed to help users who are not familiar with some of the particularities of the \Planck\ products, and can be categorized into distinct groups: map operations including component subtraction, unit conversion, colour correction, bandpass transformation, and masking of map-cutouts/full-sky maps; component separation codes, map-making codes and effective beam-averaging. In addition, the PLA includes an interface to the latest version of the \planckskymodel\ simulation
tool,\footnote{\url{https://pla.esac.esa.int/\#plaavi_psm}}
with a simple user interface that allows users to simulate the microwave/sub-mm sky with \Planck, as well as future CMB experiments and custom-defined instruments.

In the coming years it is planned to continue releasing data products
from improved reprocessing of \Planck\ data and
combining \Planck\ with other experiments.

\subsection{High angular resolution SZ observations with MISTRAL}\label{subsect:SZ_Mistral}

High angular resolution mm observations are key to understand a wide variety of scientific cases. 
Among the others, 
the interaction of CMB photons with the hot electron gas in GCs and surrounding medium promises to study GCs and their deviation from relaxed behaviour. When the CMB photons scatter off a hot electron gas, for example in GCs, in fact, they undergo inverse Compton scattering which is visible in the frequency spectrum of the CMB. The resulting distorted spectrum is 
\begin{equation}
\frac{\Delta I(x)}{I_0} = y \frac{x^4e^x}{(e^x-1)^2}\left(x\coth \frac{x}{2}-4 \right)= yg(x) \,,
\end{equation}

\noindent
where
$I_0 = (2h_P/c^2)(k_B T_{CMB}/h_P)^3$, 
$T_{CMB}$ is the CMB temperature,
$x=h_P\nu/(k_{B}T_{CMB})$ is the dimensionless frequency, and
$y=\int{n_{e}\sigma_{T} k_{B}T_{e} / (m_{e}c^{2})}dl$ is the Comptonization parameter, the integral 
along the line of sight
of the electron density $n_{e}$ multiplied by the electron temperature $T_{e}$;
$\sigma_T$ is cross-section of Thomson scattering, $k_{B}$ and $h_P$ are the Boltzmann and Planck constants.
This is the tSZ effect\cite{SZ_1972,Birkinshaw_1999} and it can be used to study both relaxed and non-relaxed GCs. In fact, GCs can experience a wide variety of situation including collisions, merging, and non-relaxed situation. In addition, not only GCs can host relativistic electrons.  Hydrodynamical simulations \cite{Cen_1999,Tuominen_2021} suggested that GCs occupy the knots of the so called cosmic web (CW) and WHIM is disposed connecting GCs forming filaments that could be seen through the SZ effect itself.

An instrument being prepared for such scientific cases is the MIllimeter Sardinia radio Telescope Receiver based on Array of Lumped elements kids\cite{EBfullGrossmanpaper} (MISTRAL), to be fielded at the 64m
Sardinia Radio Telescope\footnote{\url{http://www.srt.inaf.it}} (SRT). MISTRAL will use an array of 408 kinetic inductance detectors (KIDs) with a field of view of
$4^{\prime}$
and
an angular resolution of 
$\sim 12^{\prime\prime}$.
It is a cryogenic instrument with detectors cooled down at 270\,mK and with frequency domain multiplexing (FDM), 
Reconfigurable Open Architecture Computing Hardware 2 (ROACH2) based read-out.

\begin{figure}[t]
    \centering
        \includegraphics[width=55mm]{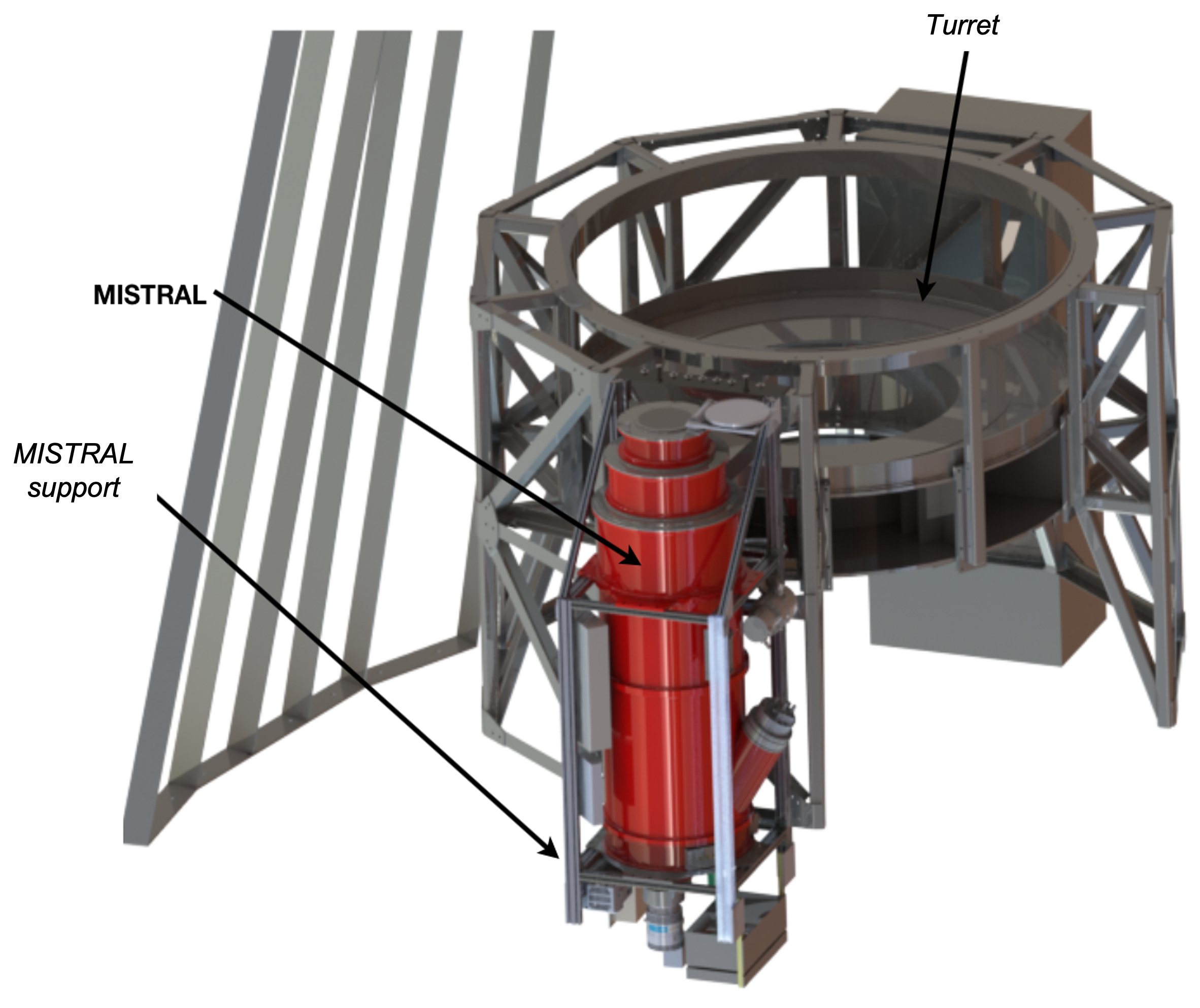}
     \includegraphics[width=30mm]{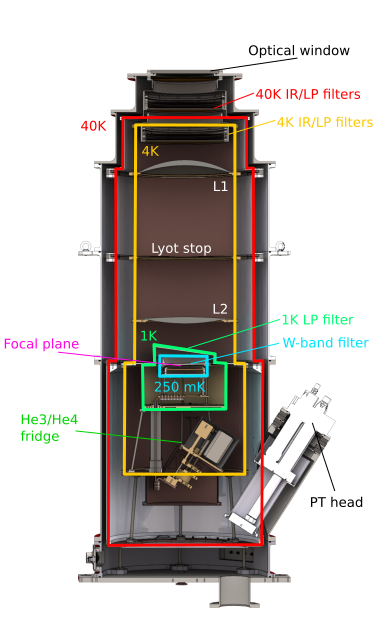}
                 \vskip -0.15cm
    \caption{{\it Left}: SRT Turret with MISTRAL installed. {\it Right}: a cut of MISTRAL cryostat highlighting the main parts. Acknowledgement: MISTRAL Collaboration.}
    \label{fig:cryo}
         \vskip -0.2cm
\end{figure}

MISTRAL is a facility instrument to be installed at the SRT in 2022. It operates in an atmospheric window in the frequency range 78$\div$103\,GHz, namely W-band, that is interesting for a number of scientific reasons, and most importantly it provides low optical depth and allows high efficiency observations. It will be installed in the Gregorian room of the SRT (Fig. \ref{fig:cryo}).

MISTRAL consists of a cryostat, being constructed by Queen Mary College (QMC) Instruments,\footnote{\url{http://www.terahertz.co.uk/qmc-instruments-ltd}} with a Pulse Tube (PT) cryocooler, an He-10 sorption refrigerator, custom optics and detectors. In order to have the PT heads as close as possible to vertical position during observations, we have positioned the head with an inclination of 57.5$^{\circ}$ with respect to the focal plane. This allows observation elevations in the range of 32.5$^{\circ}$$\div$82.5$^{\circ}$ with no degradation of the thermal performance.  

The optical design of MISTRAL includes a set of radiation quasi-optical filters, anchored at the different thermal stages of the cryostat, an Anti Reflection Coated (ARC) Ultra High Molecular Weight (UHMW) polyetilene window, and two Silicon lenses able to image the Gregorian Focus on the array of detectors. Detectors are coupled to radiation through open space (filled array) so cryogenic cold stop, placed in between the two lenses, is needed.

MISTRAL will take advantage of the high sensitivity and 
capability of FDM of KIDs cryogenic detectors. KIDs are superconductive detectors where mm radiation with higher energy with respect to the Cooper pair binding energy, can break Cooper pairs producing a change in the population densities and thus in the kinetic inductance. In fact, the inductance $L$ of a thin superconductor is dominated by the kinetic inductance $L_k$, which depends on the Cooper pair density. The detector array is composed of $408$ KIDs detectors. They are 3\,mm\,$\times$\,3\,mm each and are arranged on an equilateral triangle every 4.2\,mm on a 4\,inches silicon wafer. They sample the focal plane with a full width half maximum (FWHM) angular spacing of
$10.6^{\prime\prime}$
lower the each pixel angular resolution.
The ROACH2 based FDM,
originally developed for the 
Osservatorio nel Lontano Infrarosso Montato su Pallone
Orientabile\footnote{Far Infrared Observatory Mounted on Orientable Balloon. See \url{http://olimpo.roma1.infn.it.}}
(OLIMPO) experiment,\cite{Paiella_2019_olimpo,Masi2019} will be used.

MISTRAL will be a facility instrument:
it will be open to the scientific community to decide what kind of scientific output it can achieve and propose the observations to the Time Allocation Committee of the 
SRT.
Nevertheless, we will list some scientific cases below, including those we find most interesting.

Protoplanetary discs mm 
measurements in star forming regions allow to break the degeneracy present in infrared (IR) measurements due to the optically thick nature of the hot inner disc.\cite{Petersen_2019} 
Observations of star formation in molecular clouds with high enough angular resolution allow to distinguish starless cores with respect to those hosting protostars.\cite{Sokol} 
Continuum resolved galaxies observations can give information about morphology and radial profiles, e.g., gas column profiles, dust temperature profiles.\cite{Wall}
Spatially resolved high-$z$ radio galaxies can provide information about cold dust re-emission.\cite{Humphrey}

What is probably the most impacting expected result of MISTRAL is the possibility to observe and resolve the SZ effect through GCs and surrounding medium. With respect to moderate to low angular resolution observations, high
($\simeq 10^{\prime\prime}$) 
detections can investigate non relaxed GCs, study merging GCs,
understanding the self similarities of GCs,
their identical appearance regardless of their mass or distance.
Study the pressure profiles and understand the active galactic nuclei (AGN) feedback expected in some environments.
Often GCs are assumed to be spherical and isothermal. Nevertheless, they interact, collide and merge,
bringing
to important degeneracies and deviation from specific models. High angular resolution SZ measurements allow to disentangle between models and 
to identify the most appropriate ones. 
Even relaxed GCs experience pressure fluctuations and compressions of the intra cluster medium which would be interesting to study.

GCs experience hierarchical growth through mergers. Pre-mergers GCs should be connected through the CW and simulations endorse this.\cite{Cen_1999,Tuominen_2021}
WHIM are expected to be distributed as over-densities in filamentary structures between GCs: high angular resolution SZ measurements can detect WHIM better than X-ray under low density circumstances. This is probably the most challenging and most rewarding of the achievements that an experiment like MISTRAL can achieve.

\section{\textit{Planck} cosmological implications}\label{sect:PlanckCosmo}

\subsection{New \textit{Planck} thermal SZ map and its cosmological analysis}\label{subsect:PlanckSZMap}

The $\Lambda$CDM model 
provides a wonderful fit to many cosmological data. However, a slight discrepancy was found in the latest data analyses between the CMB anisotropies ($z\sim1100$) and other cosmological probes at
low ($z\sim0\div1$) redshift, $z$, for the $S_8 (\equiv \sigma_8 (\Omega_M / 0.3)^{0.5})$ cosmological parameter representing the amplitude of the structure growth in the universe; here $\Omega_M$ 
and $\sigma_8$ are the usual non-relativistic matter density parameter and the density contrast on a scale of 8 [$H_0$/(100\,km\,s$^{-1}$\,Mpc$^{-1})$]$^{-1}$ Mpc.

For example, the $S_8$ value was precisely measured to be $S_8 = 0.830 \pm 0.013$ with the \Planck\ CMB observation.\cite{Planck2020VI} However, a lower value of $S_8 \sim 0.77 \pm 0.03$ was measured using the population of GCs detected by \Planck\ at low-$z$\cite{Planck2014XX, Planck2016XXIV} ($z<0.6$), thus called ``$S_8$ tension''. Furthermore, other low-$z$ observations with gravitational lensing by the Kilo-Degree Survey\cite{Heymans2021} (KiDS) and 
DES
experiments also found lower $S_8$ values. These results indicate that the cosmic structure growth is slower than the prediction based on the CMB measurement and may demand modifications to the standard cosmological model to explain all these measurements.

     \begin{figure}[t]
    \centering
    \includegraphics[width=0.33\linewidth]{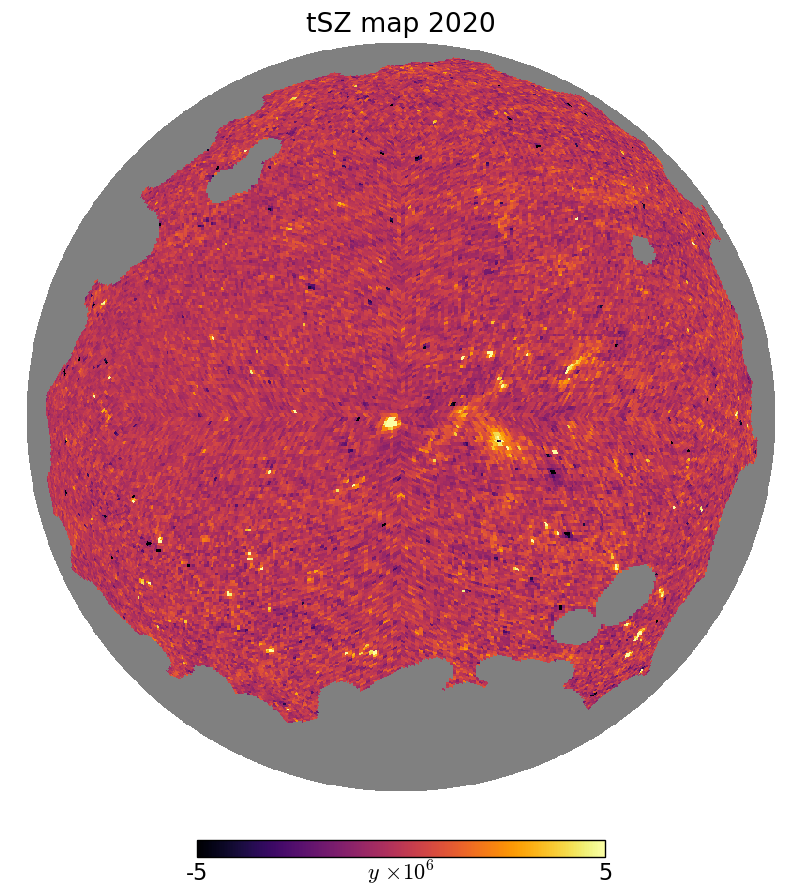}
    \includegraphics[width=0.33\linewidth]{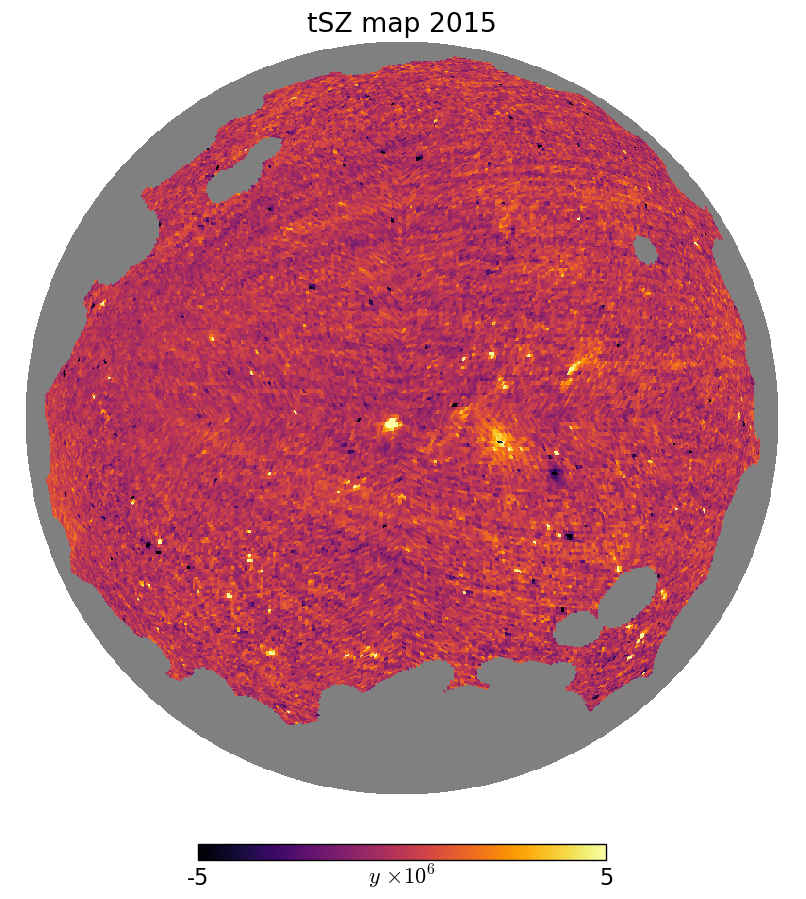}
    \includegraphics[width=0.38\linewidth]{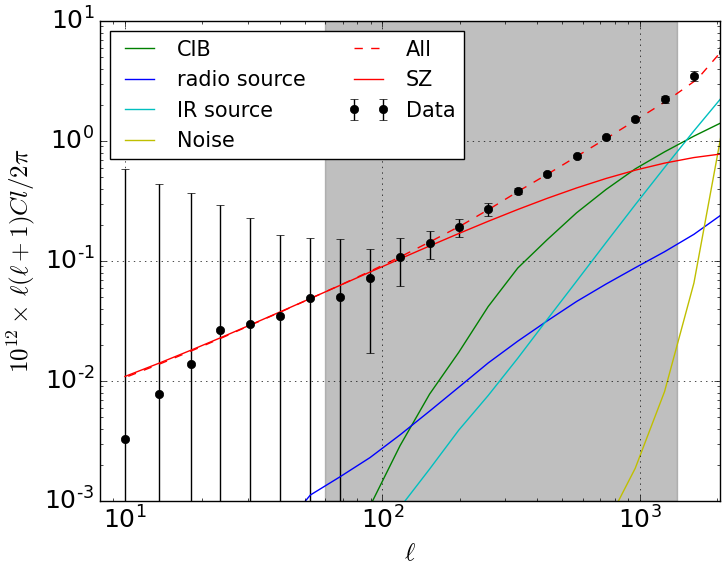}	
    \includegraphics[width=0.40\linewidth]{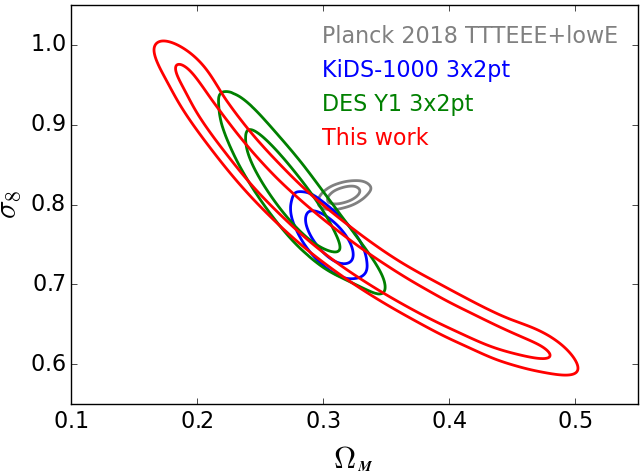}
        \vskip -0.1cm
        \caption{{\it Top}: Orthographic projections of the all-sky Compton $y$ parameter maps reconstructed with
        \milca\
        ({\it left}) and in 2015\cite{Planck2016XXII} ({\it right}), with
        appreciable differences in the striping features.
        Pixel resolution
         changed for visualization purposes from $N_{\rm side} = 2048$ to $N_{\rm side} = 128$ (i.e. from
         $\simeq 1.72^{\prime}$ to
                  $\simeq 27.48^{\prime}$ pixel size), according to the sky pixelization in Ref.~\citenum{2005ApJ...622..759G}. 
{\it Bottom}: Cross-PS (black) fitted with the tSZ (red) and foreground models,
    namely the CIB (green), radio (blue) and IR (cyan) sources, plus noise (yellow), the sum of the tSZ, foreground models and noise being shown in red dashed line ({\it left}); 
    posterior distribution of the cosmological parameters $\sigma_8$ and $\Omega_M$, with 68\% and 95\% confidence level (CL) 
    interval contours obtained from our cosmological analysis ({\it right}),
    compared with the \Planck\ CMB's (gray), KiDS-1000 3x2pt (blue), and DES Y1 3x2pt results (green).
    Acknowledgements: H. Tanimura, M. Douspis, N. Aghanim (CNRS/Univ. Paris-Saclay).}
    \label{fig:ymap}
     \vskip -0.2cm
    \end{figure}

One of the most interesting signals extensively studied with \Planck\ is the tSZ effect, 
which produces what are known as $y$ distortions, or $y$ type signals.
These primarily come from CMB photons being inverse-Compton scattered through hot GCs (see Sect. \ref{subsect:SZ_Mistral}), which makes holes,
or lowers the flux of low frequency photons and up scatters, or makes an excess flux of high frequency photons. 
 
The tSZ signal is subdominant relative to the CMB and other foreground emissions in the \Planck\ bands. 
Thus a tailored component separation algorithm is required to reconstruct the tSZ map. We adopted the 
Modified Internal Linear Combination Algorithm\cite{Hurier2013} 
(\milca)
used for the \Planck\ $y$-map reconstruction in 2015 and applied it to the 100 to 857 GHz frequency channel maps (see Sect. \ref{subsect:PLA})
from the fourth \Planck\ Release (PR4).

The \Planck\ PR4 data implemented several improvements from the previous version: the usage of foreground polarization priors during the calibration stage to break scanning-induced degeneracies, the correction of bandpass mismatch at all frequencies, the inclusion of 8\% of data collected during repointing maneuvers, etc.
With these improvements we produced a new $y$-map with smaller noises (by $\sim$7\%) and reduced
strips than the previous version released in 2015
(Fig.~\ref{fig:ymap}, {\it top} panels).

We also reconstructed two $y$-maps from the first and last halves of the data for a cosmological analysis
and considered a cross-angular PS between these $y$-maps to avoid the bias induced by the noise in the auto-angular PS (see Ref. \citenum{HTfullGrossmanpaper} for more details).
This tSZ cross-PS is affected by residual foreground emissions from radio and IR point sources and CIB emission. 
Thus we fitted our measurement with the cosmology-dependent tSZ model, including the radio and IR point source models,\cite{Delabrouille2013} the CIB model\cite{Maniyar2021} and the noise
(Fig.~\ref{fig:ymap}, {\it left bottom} panel). 

The cosmological analysis of the tSZ cross-PS allowed us to set constraints on cosmological parameters, mainly of the $S_8$ parameter. We obtained $S_8 = 0.764 \, _{-0.018}^{+0.015} \, (statistical) \, _{-0.016}^{+0.031} \, (systematic) $, in which the systematic uncertainty includes the contributions from the mass bias and pressure profile model. Our obtained $S_8$ value is fully consistent with recent weak lensing results from KiDS and DES. It is also consistent with the \Planck\ CMB's result\cite{Planck2020VI} within 2$\sigma$, while it is slightly lower by $\sim$1.7$\sigma$
(Fig.~\ref{fig:ymap}, {\it right bottom} panel).

\subsection{The CMB dipole and the thermal SZ effect: Eppur Si Muove}\label{subsect:CMBdipole}

The largest temperature anisotropy in the CMB is the dipole pattern at $\ell = 1$. The simplest interpretation of the dipole is that it is due to our motion with respect to the rest frame of the CMB (with debate over the possibility of alternative explanations), and is measured to be $\beta=v/c=(1.23357\pm0.00036)\times10^{-3}$ in the direction of the constellation Crater,
$(l,b)=(264.021^{\circ}\pm0.011^{\circ},48.253^{\circ}\pm0.005^{\circ})$,
see Ref.~\citenum{planck2016-l01}. As well as creating the $\ell$=1 mode of the CMB sky, this motion affects all astrophysical observations by modulating and aberrating sources across the sky, such as the 
CMB temperature anisotropies.
These and other effects will be discussed. 

As discussed above, an accurate mapping of the tSZ signal on the sky has been achieved with the \Planck\ satellite.
This allows us to construct a novel and independent measure of the CMB dipole because the $y$ maps are contaminated by temperature anisotropies stemming from the CMB dipole. This is also valuable as a test of the quality of the $y$ maps.

To derive the connection between the $y$ map and the dipole we will begin by defining some useful terms regarding the unboosted CMB sky,
\begin{align}
        I(x)&\equiv\frac{2k^3_\mathrm{B}T_{CMB}^3}{h_P^2c^2}\frac{x^3}{e^x-1} \\
    f(x)&\equiv\frac{xe^x}{e^x-1} \\
    Y(x)&\equiv x\frac{e^x+1}{e^x-1}-4 \,,
\end{align}

\noindent
which are 
the standard Planck blackbody intensity function, the frequency dependence of
CMB anisotropies, and the relative frequency dependence of the tSZ effect or $y$ type distortions,
$x$ being the dimensionless frequency (see Sect. \ref{subsect:SZ_Mistral}).
 Thus, to first-order the anisotropies of intensity measured by \Planck\ can be written as
 \begin{align}
     \frac{\delta I(\hat{\vec{n}})}{If(x)}=\frac{\delta T(\hat{\vec{n}})}{T_{CMB}}+y(\hat{\vec{n}})Y(x) \,,
     \label{equ:CMBsky}
 \end{align}

\noindent
where we have included just the temperature anisotropies and the $y$ signals here.

If we apply a boost to Eq.~(\ref{equ:CMBsky}) with
a dimensionless velocity $\bm{\beta}$, 
it becomes
\begin{align}
    \frac{\delta I'(\vec{\hat{n}'})}{If(x)}=&\beta\mu+\frac{\delta T(\vec{\hat{n}'})}{T_{CMB}}(1+3\beta\mu)\nonumber\\
    &+Y(x)\left(y(\vec{\hat{n}'})(1+3\beta\mu)+\beta\mu\frac{\delta T(\vec{\hat{n}'})}{T_{CMB}}\right)\nonumber\\
    &+\beta\mu y(\vec{\hat{n}}')\left(Y^2(x)-x\frac{dY(x)}{dx}\right)+\mathcal{O}(\beta^2)
    \label{equ:boostCMB} \,;
\end{align}
here $\mu={\rm cos}(\theta)$, and $\theta$ is the angle from the direction of the boost $\bm{\beta}$ and the line of sight.
Additionally, the aberration effect deflects the direction of the incoming photons from $\vec{\hat{n}}\rightarrow \vec{\hat{n}'}$ where $\vec{\hat{n}'}=\vec{\hat{n}}-\nabla(\vec{\hat{n}}\cdot\bm{\beta})$.
The first line has the same frequency dependence as thermal fluctuations. 
The term $\beta\mu$ simply describes the pure CMB dipole, as discussed previously. The second term represents the dipole aberration and modulation of the temperature anisotropies of the CMB. This was first measured in Ref.~\citenum{planck2013-pipaberration} to
$5\sigma$.
Crucially for our analysis, the middle line has the same frequency dependence as $y$ type distortions, and so encompasses the signals in the $y$-map. The final line has more obscure frequency dependencies and so will not be discussed here.
See Refs.~\citenum{planck2020-LVI,RSfullGrossmanpaper} for further details.

The second line of Eq.~(\ref{equ:boostCMB}) shows the signals in the $y$ maps produced by \Planck. The first half is the same boost as was seen in the temperature anisotropies, however, the final term is due to the second-order expansion of the intensity about $T_{CMB}$ and adds a contribution to the $y$ maps from the temperature anisotropies. We measured this signal in Ref.~\citenum{planck2020-LVI} to $6\sigma$ by cross-correlating the expected modulated temperature map with the $y$ maps. We confirmed the dipole direction by cross-correlating the temperature maps modulated in two additional perpendicular directions. To test consistency and rule out different systematics,
we used the Internal Linear Combination (\ilc)
in two dimensions, \twodilc, and \milca\ $y$-maps, the temperature anisotropies from both \twodilc\ and 
\smica,
and finally
both a harmonic space analysis and a map space analysis. 

The question as to whether an intrinsic dipole could be observationally distinguished from an extrinsic dipole remains an open question.
The terms discussed in Eq.~(\ref{equ:boostCMB}) are based on the assumption of a 
CMB blackbody spectrum\footnote{See instead Ref.~\citenum{2021A&A...646A..75T} for a study exploiting the relaxation of the blackbody assumption and references to other methods.}
and cannot be used to distinguish the two,
as they would naturally arise whether the CMB dipole were caused by a boost, or if there were simply a dipole in the sky with the same magnitude and direction.

\subsection{\textit{Planck} constraints on the primordial tensor-to-scalar ratio}\label{subsect:Planck_r}

\newcommand{\NPIPE}{{\tt NPIPE}}
\newcommand{\lollipop}{{\tt LoLLiPoP}}
\newcommand{\hillipop}{{\tt HiLLiPoP}}

The CMB anisotropies are characterized by the four Stokes parameters $[I, Q, U, V]$, or $[T, Q, U, V]$, representing the overall radiation intensity ($I$), or temperature ($T$), 
the linear polarization ($Q$ and $U$) and the circular polarization ($V$).
Neglecting the circular 
polarization,\footnote{See Sect. \ref{subsect:LorentzViol} for a discussion of some of the perspectives based on the study of the circular polarization.} 
from the conventional Stokes parameters $Q$ and $U$ it is possible to construct two rotationally invariant fields $E$ and $B$, which are linear, non-local, 
combinations of $Q$ and $U$. 
In the limit of Gaussian fluctuations, the statistical properties of CMB anisotropies can be fully characterized by the temperature ($T$), polarization ($E$ and $B$ modes) 
and cross-correlation\footnote{For simplicity, we use here two letters only for cross modes.
{Since under reflections (parity transformations) $E$ transforms as a scalar (like $T$) while $B$ transforms as a pseudo-scalar,
in absence of special rotation effects the cross-correlations $TB$ and $EB$ vanish.}}
$TE$ angular PS, $C_\ell^{X}$, where $X$ stands for a given
mode.\cite{1997ApJ...482....6S,1997PhRvL..78.2058K,1997PhRvL..78.2054S}
$C_\ell^{X}$ is typically represented in terms of $D_\ell^X = \ell (\ell+1) C_\ell^X / (2\pi)$, which is nearly flat at low multipoles for $X=T$,
although a power suppression at very large scales, first observed by the
Cosmic Background Explorer (COBE) and then confirmed by the Wilkinson Microwave Anisotropy Probe (WMAP) and by the \planck\ satellite,
is predicted in several inflationary models possibly
in connection with universe geometry and topology.

Primordial $B$-mode polarization (see, e.g., Ref. \citenum{HuWhite97} for an introduction to CMB polarization) 
can be generated by a variety of sources, like primordial magnetic
fields (see, e.g., Ref. \citenum{2007NewAR..51..275D} for a review and Refs. \citenum{2014PhRvL.112s1303B,2016A&A...594A..19P} for analyses from recent CMB data),
quantum fluctuations from various inflation models
(see, e.g., Refs. \citenum{1981JETPL..33..532M,1982JETP...56..258M,1982PhLB..115..295H,1982PhRvL..49.1110G,1982PhLB..117..175S,1983PhRvD..28..679B,1985JETPL..41..493M})
that predict cosmological perturbations in terms of tensor component (gravitational waves) and scalar component (density variations),
and topological defects (see, e.g., Refs. \citenum{2007PhRvD..76d3005B,2011PhLB..695...26G,2014PhRvD..89h3512F}) that, remarkably, produce $B$-modes also through vector perturbations.
The $B$-mode shape and amplitude depend on the considered model, inflationary scenarios typically predicting much less power than topological defects.

Discovering the stochastic field of gravitational waves generated during the early phases of the universe is one of the most ambitious goals of modern cosmology.

Inflationary gravitational waves entering the horizon since recombination epoch generate a tensor component
contributing to temperature, polarization and cross-correlation ($TE$) angular PS, in particular at large angular scale (or at low multipoles, $\ell \lsim 150$).
While the $E$ and $TE$ modes contain both scalar and tensor signals mainly coming from the epochs of recombination and reionization, 
the model does not predict primordial scalar fluctuations in the $B$-mode.
The detection of the primordial $B$-mode then constitutes a direct way to unravel tensor perturbations from the early universe,
and the efforts to measure of the ratio, $r$,
between the amplitudes of primordial tensor and scalar perturbations is the main scientific aim 
of many on-going and future CMB projects.

When CMB photons pass through gravitational potentials produced by cosmic structures, 
part of the $E$-mode power is transformed to $B$-mode power, the so-called lensing $B$-mode. This signal,
already detected by a number of ground-based observatories and by \Planck, is particularly important 
at small angular scale\cite{1998PhRvD..58b3003Z} ($\ell \gsim 150$), 
but it may mask the primordial $B$-mode 
also at intermediate angular scale and, for low $r$, even at large angular scale.

Assuming the spectra of scalar and tensor perturbations are described by pure power-laws, 
$\mathcal{P}_{\rm s}(k) = A_{\rm s} ({k}/{k_0})^{n_{\rm s}-1}$, $\mathcal{P}_{\rm t}(k) = A_{\rm t} ({k}/{k_0})^{n_{\rm t}}$,	
they are fully defined by the two amplitude parameters $A_{\rm s}, A_{\rm t}$ and by the so-called (comoving wavenumber) pivot scale $k_0$,
the definition of $r = A_{\rm t} / A_{\rm s}$ being then related to the choice of $k_0$. Among the various choices adopted in the literature,
$k_0 = 0.05\,\Mpc^{-1}$ approximately corresponds to the middle of the logarithmic range of $k$ probed by \planck.
For historical reasons, a scale-invariant spectrum corresponds to $n_{\rm s} = 1$ or to $n_{\rm t} = 0$ for scalar or tensor perturbations, respectively.

The multipole range $2 \lsim \ell \lsim 150$ is the most advantageous to directly extract $r$ from $B$-mode analyses. According to theoretical predictions,
the primordial $D_\ell^B$ is close to the maximum at $50 \lsim \ell \lsim 150$, the region of the so-called recombination bump (i);
at large scale, $\ell \lsim 10$, $D_\ell^B$ exhibits a pronounced bump generated at late epochs by the reionization process (ii); 
at $10 \lsim \ell \lsim 50$, ${\rm log} (D_\ell^B)$ increases almost linearly with ${\rm log} (\ell)$ (iii).
Exploiting higher $\ell$'s is of particular relevance for a very accurate treatment (delensing) of lensing $B$-mode.
$C_\ell^B$ increases almost linearly with $r$, the reionization bump features depending on $r$ and on the details of the reionization process,
which is parametrized to first approximation by the corresponding Thomson optical depth, $\tau$.
In general, the detection of primordial $B$-modes requires exquisite experimental sensitivity and accuracy as well as very precise treatment
of the polarized foreground emissions
(see, e.g., Ref. \citenum{2016A&A...586A.133P}),
particularly for low values of $r$, which is why a primordial $B$-mode tensor signal has not yet been detected.
On the other hand, recent CMB projects set significant upper limits to $r$.
The great sensitivity achievable by ground-based experiments on selected sky areas has been used to strongly constrain (i), 
the current tightest $B$-mode limits on $r$ from these scales coming from the Background Imaging of Cosmic Extragalactic Polarization (BICEP)/Keck 
measurements\cite{2018PhRvL.121v1301B} (BK15),
while 
the \planck\ all-sky survey allows to constrain (i), (ii) and (iii). In particular, Ref. \citenum{2020A&A...641A..10P} presents $B$-mode limits on $r$ 
from \planck\ data at $\ell \lsim 30$. 

A recent reanalysis of \Planck\ data alone and in combination with BK15 (see Ref. \citenum{2015PhRvL.114j1301B} for a previous joint analysis) 
has been performed in Ref. \citenum{2021A&A...647A.128T}
exploiting the whole multipole range $2 \le \ell \le 150$. In this study, the \Planck\ PR4 data (see Sect. \ref{subsect:PLA}) were adopted,
where the \NPIPE\cite{2020A&A...643A..42P} processing pipeline was used to create calibrated frequency maps in temperature and polarization from LFI and HFI data, 
providing several improvements in noise and systematics levels in both frequency and component-separated maps.
Various sky masks were considered in order to retain different sky fractions (from 30\% to 70\%) and to test the impact of data treatment.
In polarization, the CMB sky was separated from foregrounds applying the \commander\cite{2008ApJ...676...10E} code to a model with three components, namely
the CMB, synchrotron, and thermal dust emission, starting from the PLA PR4 maps but downgraded from $N_{\rm side} = 2048$ to $N_{\rm side} = 16$ and to $N_{\rm side} = 1024$
respectively for the analysis at $2 \le \ell \le 35$ and at  $35 \le \ell \le 150$.
An updated version of the LOw-$\ell$ LIkelihood on POlarised Power-spectra (\lollipop) code previously employed by the \Planck\ Collaboration in the reionization analysis\cite{2016A&A...596A.108P}
was used to derive the likelihood from {\it cross}-power spectra for the CMB maps reprocessed as outlined above.
Here, {\it cross} denotes that the angular PS is extracted analyzing different sets of detectors:
indeed, for these {\it cross}-power spectra the bias is zero when the noise is uncorrelated between maps.
Unbiased estimates of the angular PS were derived using
an extension of the quadratic maximum likelihood estimator and a classical pseudo-$C_\ell$ estimator at $2 \le \ell \le 35$ and at $35 \le \ell \le 150$, respectively.
The $C_\ell$ covariance is deduced from end-to-end simulations and thus includes CMB sample variance, statistical noise, residuals from systematics, uncertainties from foreground subtraction,
and correlations generated by masking.
These uncertainties are propagated through the likelihood up to the level of cosmological parameters. 

Applying this analysis to the $B$-mode alone, fixing the other cosmological parameters to the fiducial \lcdm\ spectrum of \planck\ 2018 results, and
considering the cleanest 50\% fraction of the sky, the values of $r$ (for $k_0 = 0.05\,\Mpc^{-1}$) retrieved from the posterior distribution analysis 
at $2 \le \ell \le 35$ (that includes the reionization bump) or at
$35 \le \ell \le 150$ (that includes the recombination bump) are
$r_{0.05} = -0.014_{-0.111}^{+0.108}$ or $r_{0.05} = +0.069_{-0.113}^{+0.114}$, respectively.
The two multipole windows contribute almost equally to the overall \planck\ sensitivity to $r$.
Combining them gives $r_{0.05} = 0.033 \pm 0.069$, while setting also a positive value of $r$ as a physical prior implies $r_{0.05} < 0.158$ at 95\% CL. 

A tighter result, $r_{0.05} = -0.031 \pm 0.046$, and a stronger constraint, 
$r_{0.05} < 0.069$ at 95\% CL, slightly improving the BK15 95\% CL limit $r_{0.05} < 0.072$, 
is derived applying the same method but jointly considering the three polarization modes
$B$, $E$ and $EB$.\footnote{The primordial $EB$ is expected to be null in the \lcdm\ model. On the other hand, special rotation effects of the polarization field (such as birefringence)
can produce some $EB$ signal.}

These results on $r$ do not depend significantly on the choice of other \lcdm\ cosmological parameters.
In particular, the constraints on $r$ are unchanged when $r$ is retrieved together with $\tau$:
considering all polarization modes $r_{0.05} = - 0.015 \pm 0.045$ and $\tau = 0.0577 \pm 0.0056$, while 
considering only the $B$-mode $r_{0.05} = 0.025 \pm 0.064$ and $\tau$ is undetermined because of the noise. 

The \planck\ temperature data are crucial to determine the other \lcdm\ parameters and then, when combined with polarization data,
to exhaustively verify if the limits on $r$ depend on the other parameters. 
Furthermore, although \planck\ temperature data, when considered alone, are found to be about two times less sensitive than \planck\ polarization data in constraining $r$,
the tensor contribution to temperature fluctuations allows to further tight the constraints on $r$ derived from polarization data alone.
In fact, including also \planck\ temperature data in the analysis using the \planck\ public low-$\ell$ temperature-only likelihood based on the PR3 CMB map recovered from the \commander\ 
and the High-$\ell$ Likelihood on Polarised Power-spectra (\hillipop) 
at $\ell > 30$, and marginalizing over the nuisance and the other \lcdm\ cosmological parameters,
from the posterior distribution for $r$ the authors\cite{2021A&A...647A.128T} derive
the currently most stringent upper limit based on \planck\ data alone: $r_{0.05} < 0.056$ at 95\% CL.

Finally, in Ref. \citenum{2021A&A...647A.128T} the \planck\ temperature and polarization data analyzed as summarized above have been combined
with the BK15 data, assuming their mutual independence, i.e. simply multiplying their likelihood distributions, as justified given their so different sky coverage, 
obtaining $r_{0.05} < 0.044$ at 95\% CL which currently represents the most stringent upper limit.

\section{Future and alternative perspectives}\label{sect:future_alternat}

\subsection{Lorentz invariance violation and CMB polarization}\label{subsect:LorentzViol}

The standard model of cosmology can explain the CMB radiation from the recombination epoch till today. 
As discussed above, it predicts the existence of some degrees of linear polarization known
as $E$-modes and $B$-modes (see also Refs. \citenum{Kosowsky:1994cy,Seljak:1996is,Zaldarriaga:1996xe}).
The linear polarization can be induced to the CMB by means of Compton scattering of CMB photons and cosmic electrons in presence of cosmological perturbations. However, the circular polarization for the CMB is a rare process in the standard model of cosmology, so that observation of any level of circular polarization can be a hint for the existence of new physics.
  The circular polarization for the CMB can be generated through Faraday conversion in which linear polarized light propagates through a medium resulting in different indexes of refraction along the two transverse axes.\cite{Cooray:2002nm,Bavarsad:2009hm} Moreover, the CMB photons scattering from vector or fermionic DM can lead to circular polarization
  of the CMB.\cite{ModaresVamegh:2019nja,Haghighat:2019rht}
  Furthermore, circular polarization for the CMB can be induced by electron-photon Compton scattering in presence of a background field such as a magnetic field\cite{Cooray:2002nm} or non-commutativity in space and time.\cite{Bavarsad:2009hm,Tizchang:2018qpe}
   
In this study, we would like to explore the effects of Lorentz invariance violation (LIV) as  a new background in the generation of circular polarization (see Ref. \citenum{STfullGrossmanpaper} for details). 
To do that we  assume  the CMB photon and cosmic electron scatter while the particle Lorentz symmetry is violated. To evaluate the LIV effect in the Compton scattering, we focus on an effective field theory approach known as standard model extension (SME). The minimal version of the SME contains every gauge invariant and observer-covariant operator, made by all the standard model (SM) fields which violate Lorentz invariance specified with new couplings.\cite{Colladay:1996iz}
In the SME, LIV is considered in a coordinate that the speed of light is constant $c$ in all frames. 
The kinetic term in electron sector of electromagnetic
quantum electrodynamics (QED) Lagrangian is modified with $c^{\mu \nu}$ tensor as\cite{Colladay:1998fq}
\begin{eqnarray}\label{QEDSME}
\mathcal{L}_{\text{QED}}^{\text{\tiny{LIV}}}&=& \frac{1}{2}i\bar{\psi} (\gamma^\nu+c^{\mu\nu}\gamma_{\mu}) \mathcal{\overleftrightarrow{D}}_{\mu} \psi -\bar{\psi} m_e\psi \,,
\end{eqnarray}

\noindent
where $\mathcal{D}_{\mu}$ represents the covariant derivative with $u \overleftrightarrow{D}^{\mu} v\equiv u D^{\mu} v-vD^{\mu}u$ , $\psi$ is the fermion field, $\gamma_{\mu}$ indicates the Dirac gamma matrices, and the mass of electron is $m_e$. We assume $c^{\mu \nu}$ to be symmetric and traceless  tensor. The $c^{\mu\nu}$ tensor depends on the frame and it is estimated in the standard reference frame, i.e. Sun-centered, celestial equatorial frame (SCCEF).\cite{Kostelecky:1999mr}
All the available bounds on LIV parameters are given in SCCEF frame to be comparable with each other. Therefore, we also use correlation between components of laboratory frame and SCCEF given
in Ref.~\citenum{Kostelecky:1999mr} to transfer the result to SCCEF frame.

The polarization of the scattered  CMB  photon from cosmic electron in presence of LIV is characterized by four Stokes parameters:\cite{Kosowsky:1994cy} $[I, Q, U, V]$.
$I$ indicates the overall radiation intensity, $Q$ and $U$ represent linear polarization and $V$ is circular polarization of the CMB radiation satisfying the inequality $I^2 \geq Q^2+U^2+V^2$.
 The time evolution of Stokes parameters, defining the CMB polarization, can be expressed through the quantum Boltzmann equation, schematically as\cite{Kosowsky:1994cy}
 \begin{equation}
 \frac{df}{dt}=C[f] \,,
 \end{equation}
\noindent
where the left-hand side involves time derivative of Stokes parameters which includes the gravitational effects and space time structure; besides, all possible interactions appear at the right-hand side of Boltzmann equation.

After expanding the intensity and polarization of the CMB radiation in terms of multipole moments in spin-weighted basis, the Boltzmann equations regarding to electron-photon Compton scattering in presence of LIV by considering only scalar perturbation in the metric, are obtained as
\begin{eqnarray}
 &&\frac{d}{d\eta}\Delta_I^{(S)} +iK\mu \Delta_I^{(S)}+4[\dot{\psi}-iK\mu \varphi]
 =C^I_{e\gamma}  \label{Boltzmann}\nonumber\\
 &&\frac{d}{d\eta}\Delta _{P}^{\pm (S)} +iK\mu \Delta _{P}^{\pm (S)} = C^\pm_{e\gamma}- \, i \kappa_\text{LIV}^{\pm}\,\Delta _{V}^{(S)} 
 \nonumber\\&&
 \frac{d}{d\eta}\Delta _{V}^{(S)} +iK\mu \Delta _{V}^{(S)} = C^V_{e\gamma}+ \frac{i}{2}\Big[\kappa_\text{LIV}^{
 	+}\,\Delta _{P}^{-(S)}+\kappa_\text{LIV}^{-}\,\Delta _{P}^{+(S)}\Big] \,;
 \label{Boltzmann1}
 \end{eqnarray}

\noindent
here $\eta$ is the conformal time, $\psi$ and $\phi$ are the scalar perturbation corresponding to the potential of Newton
and the spatial curvature perturbation, $\mu$ is the scalar product of the wave vector
$\bold{K}$ and the 
CMB
photon
propagating direction, $\Delta _{P}^{\pm (S)}=Q^{(S)}\pm iU^{(S)}$ is a linear combination of linear polarizations, $C^I_{e\gamma}$, $C^\pm_{e\gamma}$ and $C^V_{e\gamma}$ indicate the involvement of the SM Compton scattering in $\Delta _{P}^{\pm (S)}$ and $\Delta _{V}^{ (S)}$ parameters, respectively.
In presence of LIV, the terms $ \kappa_\text{LIV}$ and $\kappa_\text{LIV}^{\pm}$ in Eq. (\ref{Boltzmann1}) are given by
 \begin{eqnarray}
 \kappa_\text{LIV} = -a(\eta)\,\frac{3}{4}\, \frac{\sigma_T }{\alpha} \frac{m_e^2}{ k_0}\frac{\bar{n}_e}{m_e},\,\,\,\kappa_\text{LIV}^{\pm}=\kappa_\text{LIV}\, c_{\{\alpha \beta\}}(\rho_Q^{\{\alpha \beta\}}\pm i \rho_U^{\{\alpha \beta\}}) \, ;
 \end{eqnarray}

\noindent
they define the Compton scattering contribution, $k_0$ being the momentum of the CMB photon.
In general, the CMB polarization at the direction $\hat{\bold{n}}$ and at the present time $\eta_0$
is derived by integrating
Eq. (\ref{Boltzmann1})
along the line of sight and summing over all Fourier modes $\bold{K}$.
The circular polarization due to LIV Compton scattering can finally be estimated in terms of  linear polarization as follows
\begin{eqnarray}
\Delta _{V}^{(S)}(\mathbf{K},\mu,\eta_0)
&\approx&\frac{1}{2}\int_0^{\eta_0} d\eta\,
\dot\eta_{e\gamma}\,e^{ix \mu -\eta_{e\gamma}}\,\,\Big[ 3\mu\Delta _{V1}^{(S)}+2i\,\rho^{\{\alpha\beta\}}_Qc_{\{\alpha\beta\}}\,\frac{\kappa_\text{LIV}}{\dot\eta_{e\gamma}}\,\Delta _{P}^{(S)}\nonumber \\&+&2\frac{\kappa_\text{LIV}^-\kappa_\text{LIV}^+}{\dot\eta_{e\gamma}^2}\Delta _{V}^{(S)}\Big] \,,
\end{eqnarray}
\noindent
where
\begin{equation}\label{DP}
\Delta _{P}^{(S)}
(\mathbf{K},\mu,\eta)=\int_0^{\eta} d\eta\,\dot\eta_{e\gamma}\,
e^{ix \mu -\eta_{e\gamma}}\,\,\Big[ {3 \over 4}(1-\mu^2)\Pi(K,\eta)\Big] \,,
\end{equation}
\begin{eqnarray}\label{eq:Stoks-timec00}
\rho_{Q}^{\{\alpha \beta\}}&=&-\Big[\epsilon^1\cdot \hat{c}\cdot\epsilon^2+\sum_{f=e,p}\frac{ m_f v_b^2}{2 k_0}( \hat{v}\cdot\epsilon^2 \hat{v}\cdot \hat{c}\cdot\epsilon^1+\hat{v}\cdot\epsilon^1 \hat{v}\cdot \hat{c}\cdot\epsilon^2)\\ \nonumber &-&\frac{v_b}{2}(\hat{v}\cdot\epsilon^2 \hat{k}\cdot \hat{c}\cdot\epsilon^1+\hat{v}\cdot\epsilon^1 \hat{k}\cdot \hat{c}\cdot\epsilon^2) \Big] \,,
\end{eqnarray}

\noindent
$\Pi=\Delta_{T2}^{(S)}+\Delta_{P2}^{(S)}+\Delta_{P0}^{(S)}$ and $\dot{\eta}_{e\gamma}=a (\eta) n_e \sigma_T$ and  $\eta_{e\gamma}=\int_\eta^{\eta_0} \dot{\eta}_{e\gamma}(\eta)d\eta$ are the differential optical depth and total optical depth.
  For a detail calculation of Eqs. (\ref{Boltzmann1})-(\ref{eq:Stoks-timec00}) see Ref.~\citenum{STfullGrossmanpaper}.
The angular PS, $C_{\ell}^V$, arisen from CMB-cosmic electron forward scattering in presence of LIV is obtained as
  \begin{eqnarray} \label{cvl:int}
	C_{\ell}^V
 	&\approx&\frac{1}{2\ell+1}\int d^3KP_{v}(\bf{K})\times\nonumber\\&\sum_m&\Big|\int d\Omega Y^*_{l\ell m}\int_0^{\eta_0} d\eta\,
 	\dot\eta_{e\gamma}\,e^{ix \mu -\eta_{e\gamma}}\,2i\,\rho^{\{\alpha\beta\}}_Qc_{\{\alpha\beta\}}\,\frac{\kappa_\text{LIV}}{\dot\eta_{e\gamma}}\,\Delta _{P}^{(S)}|^2 \,;
 \end{eqnarray}
 
 \noindent
  here $P_{v}(\bf{K})$ 
 indicates the velocity PS which can be estimated as a function of primordial scalar spectrum
 $P_{v}(\bold{K},\tau)\sim P^{(S)}_\phi(\bold{K},\tau)$.
 Therefore,
we can estimate the $C_{\ell}^V$ in terms of linearly polarized angular PS as follows:
 \begin{equation}
    C_{\ell}^V \approx (\tilde\kappa^\text{avg}_\text{LIV})^2 C_{\ell}^P \,,
 \end{equation}
 
 \noindent
 with
 \begin{equation}
 \tilde\kappa_\text{LIV}= c_{\{\alpha\beta\}}\ \rho^{\{\alpha\beta\}}_Q\frac{\kappa_\text{LIV}}{\dot\tau_{e\gamma}},~~~~~\tilde\kappa^{\text{avg}}_\text{LIV}=\frac{1}{z_{lss}}\int_0^{z_{lss}} dz~ \tilde\kappa_\text{LIV}(z)\simeq 10^{-3} \,,
 \end{equation}
 
\noindent
 where $\tilde\kappa^{\text{avg}}_\text{LIV}$ is average of  $\tilde\kappa_\text{LIV}(z)$ over redshift
 from last scattering surface, $z_{lss}\simeq 1100$, up to today, $z=0$.
 Based on recent experimental astrophysics  bounds on different components of $c^{\alpha\beta}$ tensor,\cite{Kostelecky:2008ts} we assume all component of $c^{\alpha\beta}$ tensor at the order of $10^{-15}$. Considering the experimental value for the CMB linearly polarized
 angular
 PS of the order of\cite{Planck:2015sxf}
 $C_{\ell}^P\equiv 0.1 \mu$K$^2$ 
 for $\ell<250$, we can estimate the $V$-mode
 angular
 PS arisen from Compton scattering in presence of LIV as follows:
 \begin{equation}\label{cvl-bound}
 C_{\ell}^V\, \approx\,\,0.8~{\rm nK}^2 \Bigl( {  { \frac{c^{\alpha \beta}}{10^{-15}} }  } \Bigr)^2 \,.
 \end{equation}
 
The current experimental bound on the circular polarization coming from the Cosmology Large Angular Scale Surveyor (CLASS) experiment with the 40 GHz polarimeter 
ranging from
$0.4 \mu$K$^2$ to
$13.5 \mu$K$^2$ for
$1\leq \ell \leq 120$. 
We expect that by improving the sensitivity of experiments in future, the constraint obtained in Eq. (\ref{cvl-bound}) would be comparable to experimental bounds.

\subsection{New gravitational degrees of freedom and the DM problem}\label{subsect:DM}

\newcommand{\grad}{\ensuremath{\vec{\nabla}}}                                                                                                                           
\newcommand{\GN}{\ensuremath{G_{{\rm N}}}}                                                                                                                              
\newcommand{\KB}{\ensuremath{K_{{\rm B}}}}
\newcommand{\cad}{c_{\rm ad}}
\newcommand{\kpc}{\ensuremath{{\rm kpc}}}  
\newcommand{\Qcal}{{\cal Q}}
\newcommand{\Ycal}{{\cal Y}}
\newcommand{\Gt}{\ensuremath{\tilde{G}}}                                                                                                                                
\newcommand{\Fcal}{{\cal F}}
\newcommand{\Jcal}{{\cal J}}
\newcommand{\Kcal}{{\cal K}}
\newcommand{\phib}{{\bar{\phi}}}                                             

The nature of DM is the deepest problem of modern cosmology. Assuming GR, 
DM appears to be a necessary ingredient in theoretical models  constructed for explaining  astrophysical and cosmological observations on 
 $\sim\kpc$ scales or larger. This essentially boils down to a mismatch between the observed dynamics of visible matter and its gravitational influence.
Many theories of what is DM have been proposed.\cite{Bertone:2004pz}
The DM microphysical nature,
however, is not crucial to explain
many cosmological observations.
This leads to a simple description so that the DM dynamics are effectively governed by the collisionless Boltzmann equation, coupled to gravity.

The less explored alternative is that DM is not the cause behind all these observed phenomena, but rather 
that GR (and also Newtonian gravity) breaks down at ultra-low curvatures and gravitational accelerations and a new description of gravity is necessary.
Changing the law of gravity, however, is not as arbitrary as it may seem.\cite{Clifton:2011jh}
Lovelock's theorem\cite{Lovelock:1971yv} tells us that GR is the unique 4-dimensional theory based on a local and diffeomorphism-invariant action 
where the metric is the only dynamical degree of freedom.
If a non-GR gravitational theory is then assumed to explain the DM phenomenon, it is inevitable that one or more of these assumptions must be broken,
and this generically leads to
new dynamical degrees of freedom, that is, new dynamical fields. 

Milgrom proposed\cite{Milgrom1983a,Milgrom1983b,Milgrom1983c} that DM is only an apparent phenomenon and 
that one can instead fit galactic rotation curves via a modification of the inertia/dynamics of baryons
or of the gravitational law, when accelerations become smaller than $a_0\sim 1.2\times 10^{-10}$\,m/s$^2$.
Modifying the gravitational law was further explored by Bekenstein \& Milgrom\cite{BekensteinMilgrom1984} who proposed  that when
the gradients of the potential $\Phi$ are smaller than $a_0$, non-relativistic gravity is effectively governed by
\begin{equation}
 \grad \cdot\left(  \frac{|\grad\Phi|}{a_0} \grad \Phi \right)  = 4 \pi \GN \rho \,,
\label{eq_AQUAL}
\end{equation}
where $\GN$ is the Newtonian gravitational constant, and $\rho$ the matter density.
These type of models are referred to as Modified Newtonian Dynamics (MOND).\footnote{We note that rather than modifying gravity,
the superfluid dark matter model may also lead to effective MOND behaviour on galactic scales while being compatible with $\Lambda$CDM behaviour on cosmological scales,
as argued in Ref. \citenum{Berezhiani:2015bqa}.}

MOND has enjoyed success in fitting galactic rotation curves and
reproducing the baryonic Tully-Fisher relation, while on GC scales it is found that either $a_0$ must be larger by a factor of 4$\div$5 or that 
some DM is still needed at GC cores; see Ref. \citenum{FamaeyMcGaugh2011} for a review.
As MOND is inherently non-relativistic, it is difficult to test in cosmological settings, since systems such as the CMB require a relativistic treatment.                                                                                                                              
Relativistic theories that yield MOND behavior have been proposed, the most well-known being the Tensor-Vector-Scalar (TeVeS) theory,\cite{Sanders1997,Bekenstein2004}
 making clear predictions regarding gravitational lensing and cosmology.  In cases where the CMB and matter power spectra (MPS) have been computed, 
no theory has been shown to fit \emph{all} of the cosmological data while preserving MOND phenomenology in galaxies. TeVeS in particular can be compatible with the MPS
 of large scale structure\cite{SkordisEtAl2005,DodelsonLiguori2006} but fails to fit the CMB. Moreover, it leads to a tensor mode gravitational wave (GW) speed 
different than the speed of light,\cite{SkordisZlosnik2019} in contradiction with the 
Laser Interferometer Gravitational-Wave Observatory (LIGO)/Virgo
observations of GW along with an electromagnetic counterpart.

The non-relativistic equation \eqref{eq_AQUAL} can only be an effective description, due to Lovelock's theorem. To make it into a fully fledged relativistic theory, one can introduce additional fields. 
Building on TeVeS and later developments,\cite{ZlosnikFerreiraStarkman2006b,SkordisZlosnik2019} Skordis \& Zlosnik\cite{Skordis:2020eui} proposed a simple relativistic theory which 
in addition to the metric $g_{\mu\nu}$ has a scalar field $\phi$ and unit time-like vector field $A^\mu$. 
Perturbing these fields on a Minkowski background, results in perturbations of $\phi$ and $A_\mu$ which mix with the metric perturbation via gauge transformations,
i.e. the metric potentials mix with the perturbations of  $\phi$ and $A_\mu$.
Thus,  $\phi$ and $A_\mu$ are gravitational fields  and not DM fields, 
because due to this mixing it is impossible to distinguish observationally which of the three is the cause behind the ``DM phenomenon'', in analogy to the electric/magnetic field mixing resulting in the electromagnetic field.

The theory contains a free function  $\Fcal(\Ycal,\Qcal)$ of two arguments: $\Qcal = A^\mu \nabla_\mu \phi$ and $\Ycal= q^{\mu\nu} \nabla_\mu \phi \nabla_\nu \phi$ where
$q_{\mu\nu} = g_{\mu\nu} + A_\mu A_\nu$ is the $3$-metric orthogonal to $A^\mu$.
It must obey certain conditions in order for the theory to serve its purpose and to be observationally viable.
To  recover MOND on galactic scales, it is necessary that the reduced function
$\Jcal(\Ycal) \equiv \Fcal(\Ycal,\Qcal_0) / (2 -\KB)$
where $\Qcal_0$ is a non-zero constant, has the limit $\Jcal \rightarrow \Ycal^{3/2}/a_0$ when $\Ycal \sim |\grad\varphi|^2$ is small. 
Imposing that and considering virialized systems in the weak-field approximation, leads to linear equations involving the metric potential $\Phi$ 
and a scalar perturbation $\varphi$, sourced by the (baryonic) matter density.
These contain an additional term which is not present in MOND, but looks like a mass term for the potential, $\mu^2 \Phi$, where here
$\mu^2 = 2 \Kcal_2 \Qcal_0^2 / (2-\KB)$. 
It is determined by
the other parameters of the theory:
$\Qcal_0$, appearing in as part of the free function, $\KB$, related to the coupling strength of the vector field and taking values in the range $0<\KB<2$,
and $\Kcal_2$, related to the coupling strength of the scalar field. In addition, the resulting equations depend on Milgrom's constant $a_0$.
 Thus in the quasistatic weak-field situation, at least two parameter combinations are important: $a_0$ and $\mu$, although, 
depending on how the GR limit is attained, there may be further parameters as part of the free function. Due to the presence of the mass term $\mu$, the theory departs from MOND behaviour
 at ultra-low accelerations determined by $\mu$.

Consider now setups involving the relativistic nature of the theory.
The form of the action of the theory ensures that the right amount of gravitational lensing is ensured whenever $\Phi$ correctly determines the dynamics
and the tensor mode GW speed is equal to the speed of light in all situations. 
Meanwhile, on 
Friedmann-Lema{\^i}tre-Robertson-Walker (FLRW) $\phi=\phib(t)$ while $A^\mu$ aligns trivially with the time direction so that $\Ycal\rightarrow 0$ and $\Qcal \rightarrow \dot{\phib}$ and the action reduces to 
that  of shift-symmetric K-essence\cite{Scherrer2004} and the low-energy limit of the ghost-condensate theory.\cite{ArkaniHamedEtAl2003}
Defining then the reduced function
$\Kcal(\Qcal)\equiv - \Fcal(0,\Qcal) / 2$,
we require that $\Kcal \approx \Kcal_2 \left(\Qcal - \Qcal_0 \right)^2 + \ldots$, with non-zero $\Qcal_0$ (same constants as what appear above), 
 the energy density of $\phi$ scales precisely as $a^{-3}$, $a=1/(1+z)$ being here the cosmic scale factor, i.e. like pressureless matter, plus
small corrections. The equation of state and speed of sound of $\phi$ also scale as $a^{-3}$ and depend on the constants $\Kcal_2$ and $\Qcal_0$ of the action.
In the late universe limit  these tend to zero so that the linear perturbations of $\phi$ and $A_\mu$ can be re-casted by linear transformations into 
a density contrast $\delta$ and velocity divergence $\theta$ of pressureless matter. In this limit, the field equations of $\phi$ and $A_\mu$ and their contribution to the Einstein equations,
become identical to pressureless-dust equations, i.e. as for CDM.  
This ensures that the CMB and MPS spectra calculated within this theory, become parametrically close to those from $\Lambda$CDM, leading to excellent fits to the CMB 
angular PS from \Planck\ and the MPS obtained from galaxy clustering.

To summarize, the theory is not 
a MOND extrapolation
to cosmology. Rather, it recovers MOND in quasistatic situations on galactic scales, and it tends to $\Lambda$CDM behaviour on cosmological scales.
It depends on a free function, which is, however, constrained so that these two limits are attained. Doing that, the theory depends on a minimum of
4 parameters, $\KB$, $\Qcal_0$, $\Kcal_2$ and $a_0$, 
while additional parameters could be part of the free function and be related to the GR limit or the early universe.

\subsection{Exploiting the magnification bias with high-$z$ sub-mm galaxies}\label{subsect:magnSMG}

Magnification bias is due to gravitational lensing and consists in the apparent excess in the number of high-$z$ sources (the lenses sources) close to the position of low-$z$ galaxies (the lenses). In fact, the gravitational lensing deflects the light rays of high-$z$ sources causing the stretching of the apparent sky area in the region affected by the lensing. This also causes a boost in the flux density of the high-$z$ sources, making them more likely to be detected above a given instrument flux density limit (see, e.g., Ref.~\citenum{SCH92}).
The magnification bias can be measured through the cross-correlation between the low and high-$z$ objects. Given two source samples with non-overlapping redshift distributions, the excess signal when computing their cross-correlation with respect to the random case is due to the magnification bias. Being such signal related to lensing and thus to cosmological distances and the galaxy halo characteristics, it can be used to constrain cosmological parameters.

In order to get the best cross-correlation measurements, at the moment the optimal choice is to use as high-$z$ sources (background samples) the high-$z$ sub-mm galaxies (SMGs) first of all because of their steep number counts, shown with recent observations by the {\it Herschel}\, Space Observatory\cite{PIL10}
 and the South Pole Telescope\cite{CAR11} (SPT)
 strengthening their magnification. 
 Moreover, the SMGs are faint in the optical which means that they do not get confused with the foreground lens sample, being such sample invisible at sub-mm wavelength. Lastly, the SMGs redshift are usually above $z>1$ so that their selection as background sample ensures no overlap with the foreground objects. In fact, due to their extreme magnification bias, Dunne et al.
(Ref.~\citenum{DUN20}) make a serendipitous direct observation of high-$z$ SMGs. While performing a study of gas tracers with 
Atacama Large Millimeter/submillimeter Array (ALMA) observation of galaxies targeting a statistically complete sample of twelve galaxies selected at 250 $\mu$m with $z = 0.35$, magnified SMGs appears around the position of half of them.

As shown in Refs.~\citenum{GON17,BON19}, these advantages make the SMGs the perfect background sample for tracing the
(baryonic and DM) mass density profiles of galaxies and GCs
and their time-evolution and for constraining the free parameters of a halo occupation distribution (HOD) model. 
In the adopted HOD model\cite{ZHE05} all halos above a mass $M_{min}$ host a central galaxy and those above a mass $M_1$ host satellites (galaxies in the halo)
whose number is a power-law, $(M/M_1)^\alpha$, of the halo mass $M$.
In particular, the results in Ref.~\citenum{GON17} suggests that the lenses are massive galaxies or GCs, with a minimum mass of $10^{13}M_\odot$. Later on, Bonavera et al. (Ref.~\citenum{BON19}) study with the magnification bias the mass properties of a sample of quasars whose position signpost the lens at $0.2 < z < 1.0$ obtaining a $M_{min} = 10^{13.6^{+0.9}_{-0.4}} M_\odot$. This again suggests that the lensing is actually produced by halos of the size of GCs placed close to the quasar positions.

\begin{figure}[t]
    \centering
    \includegraphics[width=0.295\linewidth]{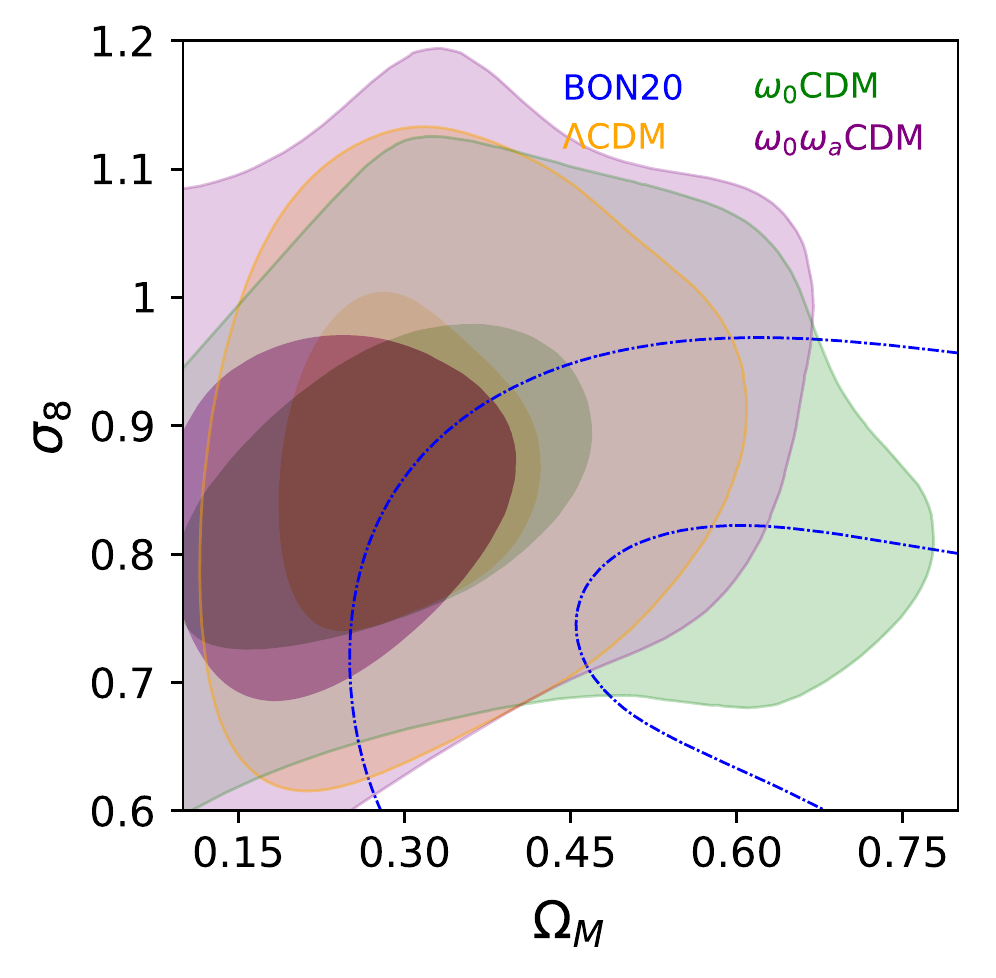}
    \includegraphics[width=0.30\linewidth]{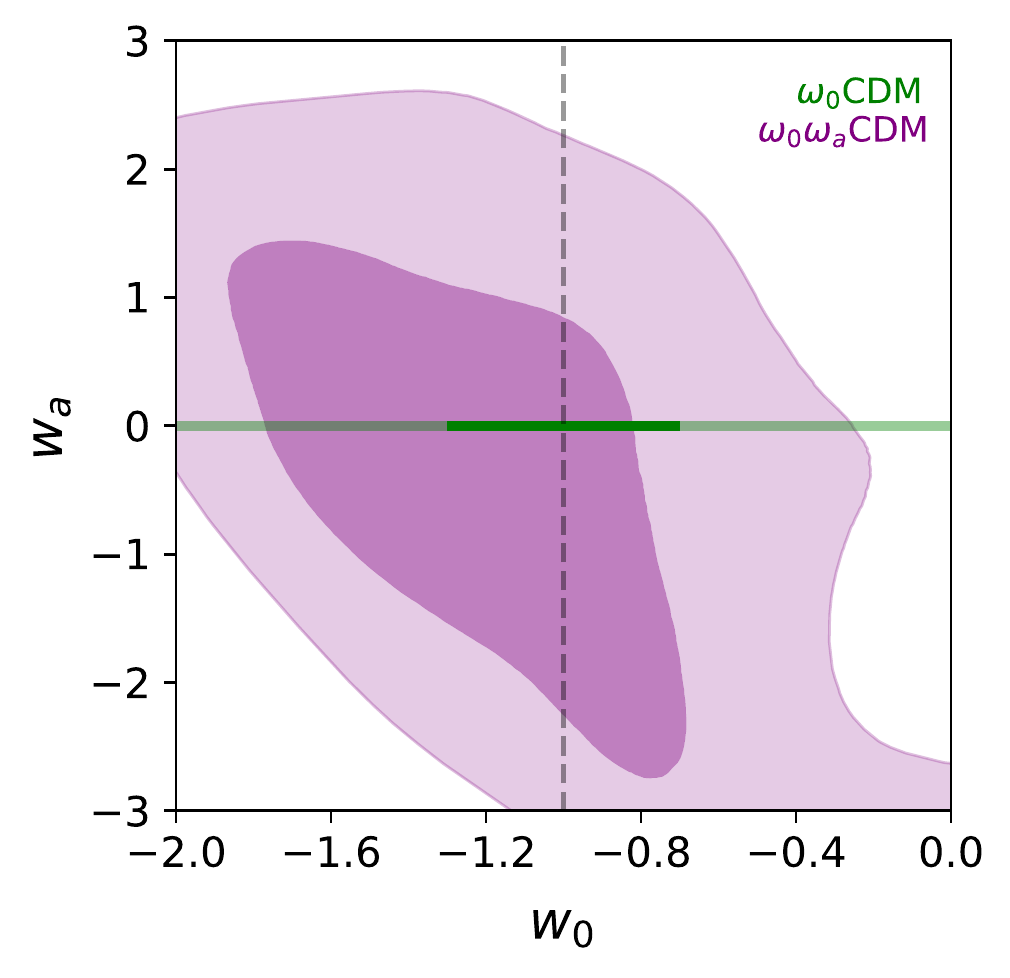}
    \includegraphics[width=0.38\linewidth]{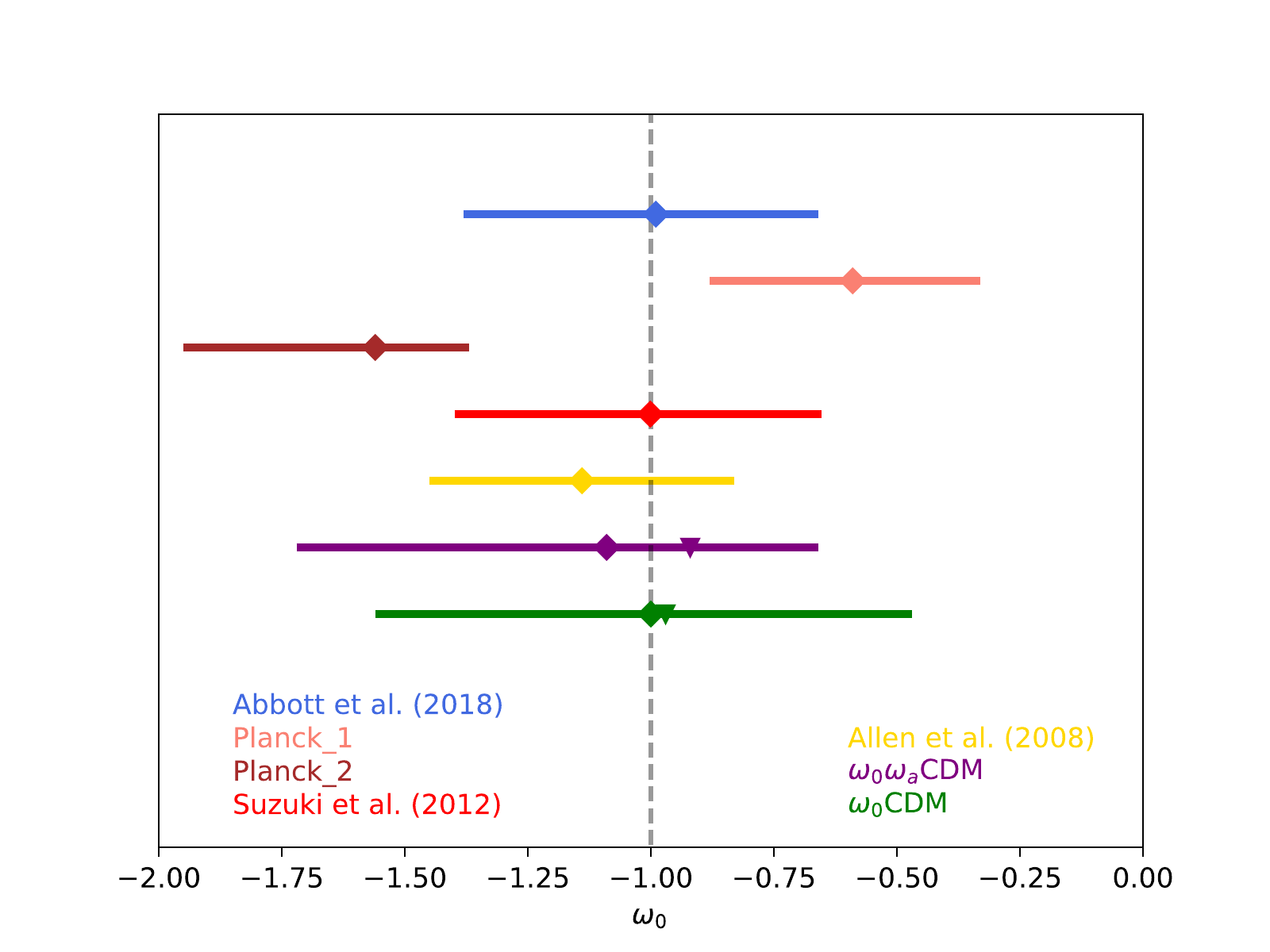}
    \vskip -0.1cm
\caption{{\it Left}:
$\Omega_M$ and $\sigma_8$ contour plot for the non-tomographic case in Ref.~\citenum{BON20} (dot-dashed blue line) and the tomographic runs in Ref.~\citenum{BON21} for $\Lambda$CDM (in yellow),
$\omega_0$CDM (in green) and $\omega_0\omega_a$CDM (in purple) models.
{\it Central}:
$\omega_0$ and $\omega_a$ contour plot for the $\omega_0\omega_a$CDM model.
In both panels, the contours of the 2-dimensional posterior distributions are set to 0.393 and 0.865. 
{\it Right}:
$\omega_0$ results for the $\omega_0$CDM (green) and $\omega_0\omega_a$CDM (purple) models compared with those by DES (blue), \textit{Planck}\_1 (base\_w\_wa\_plikHM\_TT\_lowl\_lowE\_BAO, salmon), \textit{Planck}\_2 (base\_w\_plikHM\_TT\_lowl\_lowE, brown), Supernovae (red), X-ray measurements (yellow).
Adapted from Bonavera et al., A\&A 656, A99 (2021) (Ref. \citenum{BON21}),
reproduced with permission from Astronomy \& Astrophysics, \copyright\ ESO.
}    
    \label{fig_s_O}
          \vskip 0.3cm
    \end{figure}

Moreover, some of the main cosmological parameters can be also estimated using the magnification bias, as in Ref.~\citenum{BON20}. They use the cross-correlation data measured between a foreground sample of Galaxy And Mass Assembly (GAMA) galaxies with spectroscopic redshifts in the range $0.2 < z < 0.8$ and a background sample of 
{\it Herschel} Astrophysical Terahertz Large Area Survey (H-ATLAS) galaxies with photometric redshifts $z > 1.2$ to constrain the
HOD
astrophysical parameters 
($M_{min}$, $M_1$, and $\alpha$) 
and some of the cosmological ones ($\Omega_M$, $\sigma_8$, and $H_0$). 
These parameters are estimated through a Markov chain Monte Carlo analysis.
They obtain a lower limit at 95\% CL on $\Omega_M > 0.24$, a slight trend towards $H_0 > 70$ values and a tentative peak around 0.75 with an upper limit at 95\% CL of $\sigma_8<1$. Such results are summarized in the $\Omega_M-\sigma_8$ plane in Fig.~\ref{fig_s_O} ({\it left} panel).
 
Furthermore, magnification bias trough cross-correlation measurements can be used to constrain the halo mass function. In particular, Ref.~\citenum{CUE21} successfully tests such possibility according to two common parametrizations, the Sheth \& Tormen\cite{1999MNRAS.308..119S} and the Tinker et al.\cite{2008ApJ...688..709T} fits.

They find general agreement with traditional values for the involved parameters, with a slight difference in the Sheth \& Tormen fit for intermediate and high masses, where the results suggest a hint at a somewhat higher number of halos.
Given the large number of studies that can be performed with magnification bias through cross-correlation data, a systematic analysis of possible bias that might affect the estimation is being carried out. In particular, Gonz{\'a}lez-Nuevo et al. (Ref.~\citenum{GON21}) take into account different biases at cosmological scales in the source samples to carefully measure unbiased cross-correlation functions. More precisely, their background sample consists of H-ATLAS galaxies with $z > 1.2$ whereas they use two independent foreground samples with $0.2 < z < 0.8$: GAMA galaxies with spectroscopic redshifts and 
Sloan Digital Sky Survey (SDSS) galaxies with photometric redshifts. These independent samples allowed them to perform a pseudo-tomographic study that yields to constrain
$\Omega_M=0.50^{+0.14}_{-0.20}$ and $\sigma_8 =0.75^{+0.07}_{-0.10}$. Such analysis also suggests that a tomographic approach might improve the results.

Driven by the conclusions mentioned above, 
adopting the unbiased sample by Ref.~\citenum{GON21}, 
Bonavera et al.\cite{BON21}
use magnification bias in tomography to jointly constrain the HOD astrophysical parameters 
in each selected redshift bin
together with the $\Omega_M$, $\sigma_8$, and $H_0$ cosmological ones in the $\Lambda$CDM scenario and 
for DE models in the $\omega_0$CDM and $\omega_0\omega_a$CDM frameworks
characterized by the $\omega_0$ and $\omega_a$ parameters
defining the DE barotropic index as $\omega(a)=\omega_0+\omega_a (1-a)$. 
The foreground sample has been divided into four redshift bins
(0.1--0.2, 0.2--0.3, 0.3--0.5 and 0.5--0.8)
and the sample of H-ATLAS galaxies has photometric redshifts $z > 1.2$. 
As for the HOD parameters, $M_{min}$ shows a trend towards higher values at higher $z$ confirming the findings in Ref.~\citenum{GON17}. 
For the $\Lambda$CDM model, they obtain a maximum posterior value [68\% CL] of 
$\Omega_M$ = 0.26 [0.17,0.41] and of $\sigma_8=0.87$ [0.75,1]
(implying $S_8 \simeq 0.81$, in good agreement with the result in Sect. \ref{subsect:PlanckSZMap}), 
being $H_0$ not yet constrained.
For the $\omega_0$CDM model, they find similar results on $\Omega_M$ and $\sigma_8$ and a maximum posterior value [68\% CL] of $\omega_0=-1$ [$-1.56, -0.47$]. For the $\omega_0\omega_a$CDM model, $\omega_0=-1.09$ [$-1.72,-0.66$] and $\omega_a=-0.19$ [$-1.88, 1.48$].
Fig.~\ref{fig_s_O} shows the results in the $\Omega_M-\sigma_8$ plane for the
$\Lambda$CDM, 
$\omega_0$CDM
and $\omega_0\omega_a$CDM 
models ({\it left} panel) and in the $\omega_0-\omega_a$ plane ({\it central} panel).
The results on $\omega_0$
for the $\omega_0$CDM 
and $\omega_0\omega_a$CDM 
cases are shown in Fig.~\ref{fig_s_O} ({\it right} panel), together with a comparison with other results from literature. 
Moreover, the tomographic analysis presented in Ref.~\citenum{BON21} confirms that magnification bias results do not show the degeneracy found with cosmic shear measurements and that, related to DE,
do show a trend of higher $\omega_0$ values for lower $H_0$ values (see Ref. \citenum{LBfullGrossmanpaper} for details). 

In conclusion, SMGs turn out to be
a perfect background sample for magnification bias analyses, useful in astrophysical studies and as a further independent cosmological probe for $\Lambda$CDM and beyond $\Lambda$CDM models
and this can be implemented adopting different foreground samples.
Up to now $\Lambda$CDM compatible results have been found with tomographic and non-tomographic approaches
and serendipitous direct observation of magnified SMGs have been performed with ALMA.

{
\section*{Acknowledgments}

CB acknowledges partial support from the INAF PRIN SKA/CTA project FORECaST.
LB acknowledges the PGC 2018 project PGC2018-101948-B-I00 (MICINN/FEDER).
TRC acknowledges support of the Department of Atomic Energy, Government of India, under project no. 12-R\&D-TFR-5.02-0700.
CS acknowledges support from the European Research Council under the European Union's
Seventh Framework Programme (FP7/2007-2013) / ERC Grant Agreement n. 617656 ``Theories
and Models of the Dark Sector: Dark Matter, Dark Energy and Gravity''  and from the European Structural and Investment Funds
and the Czech Ministry of Education, Youth and Sports (MSMT) (Project CoGraDS - CZ.02.1.01/0.0/0.0/15003/0000437).
RS acknowledges the support of the Natural Sciences and Engineering Research Council of Canada.
HT acknowledges support from the ERC-2015-AdG 695561 (ByoPiC project).
AW acknowledges funding from the South African Research Chairs Initiative of the Department of Science and Technology and the National Research Foundation of South Africa. 
Some of the results in this paper have been derived using the \healpix\cite{2005ApJ...622..759G} package.
The use of the PLA is acknowledged.
It is a pleasure to thank 
N. Aghanim, D. Contreras, M. M. Cueli, M. Douspis, B. Gaensler, J. Gonzalez-Nuevo, R. Mohammadi,  D. Scott, S. S. Xue, 
the MISTRAL Collaboration and the {\it Planck} Collaboration 
for long-standing collaborations on the various topics discussed in this paper.
We also thank the anonymous referee for comments that helped improve the paper.
}

{

\begin{thebibliography}{100}

\bibitem{DiValentino:2021izs}
E.~{Di Valentino}, O.~{Mena}, S.~{Pan} {\em et~al.},
{\em Classical and Quantum Gravity} {\bf 38}, 
153001 (2021).

\bibitem{DES2018}
{Dark Energy Survey Collaboration}, {\em Phys. Rev. D} {\bf 98}, 
043526
(2018).

\bibitem{DES2021}
{DES Collaboration}, 
arXiv:2105.13549
(2021).

\bibitem{1978AnPhy.115...78B}
R. Brout, F. {Englert} and E. {Gunzig},
  {\em Annals of Physics} {\bf 115}, 
  78
  (1978).

\bibitem{1980PhLB...91...99S}
A.~A. Starobinsky, 
  {\em Physics Letters B} {\bf 91},
    99
  (1980).

\bibitem{1980ApJ...241L..59K}
D. Kazanas, 
  {\em \apjl} {\bf 241},
    L59
  (1980).

\bibitem{1981PhRvD..23..347G}
A.~H. Guth, 
  {\em \prd} {\bf 23},
  347
  (1981).

\bibitem{1981MNRAS.195..467S}
K. Sato, 
  {\em MNRAS} {\bf 195},
  467
  (1981).

\bibitem{1982PhLB..108..389L}
A.~D. Linde, 
  {\em Physics Letters B} {\bf 108}, 
  389
  (1982).

\bibitem{1982PhRvL..48.1220A}
A. Albrecht and P.~J. Steinhardt,
  {\em \prl} {\bf 48}, 
  1220
  (1982).

\bibitem{1983PhLB..129..177L}
A.~D. Linde, 
  {\em Physics Letters B} {\bf 129}, 
  177
  (1983).

\bibitem{SZ_1970a}
R.~A. Sunyaev and Y.~B. Zel'dovich, 
  {\em Astrophysics and Space Science} {\bf 7}, 
  3
  (1970).

\bibitem{SZ_1970b}
R.~A. Sunyaev and Y.~B. Zel'dovich, 
  {\em Astrophysics and Space Science} {\bf 7}, 
  20
  (1970).

\bibitem{Harrison_1970}
E.~R. Harrison,
{\em Phys. Rev. D} {\bf 1}, 2726  (1970).

\bibitem{PY_1970}
P.~J.~E. Peebles and J.~T. {Yu},
{\em \apj} {\bf 162},
815
(1970).

\bibitem{Z_1972}
Y.~B. Zel'dovich, 
{\em \mnras} {\bf 160}, 1P
(1972).


\bibitem{Lorimer:2007qn}
D.~R. {Lorimer}, M.~{Bailes}, M.~A. {McLaughlin}, D.~J. {Narkevic} and F.~{Crawford},
  {\em Science} {\bf 318}, 
  777
(2007).

\bibitem{Tendulkar:2017vuq}
S.~P. {Tendulkar}, C.~G. {Bassa}, J.~M. {Cordes} {\em et~al.},
{\em \apjl} {\bf 834}, 
L7 
(2017).

\bibitem{Walters:2017afr}
A.~{Walters}, A.~{Weltman}, B.~M. {Gaensler}, Y.-Z. {Ma} and A.~{Witzemann},
{\em \apj} {\bf 856}, 
65 
(2018).

\bibitem{Walters:2019cie}
A.~{Walters}, Y.-Z. {Ma}, J.~{Sievers} and A.~{Weltman},
{\em \prd} {\bf 100}, 
103519 
(2019).

\bibitem{Weltman:2019cqv}
A.~{Weltman} and A.~{Walters}, 
arXiv:1905.07132 
(2019).

\bibitem{Platts:2018hiy}
E.~{Platts}, A.~{Weltman}, A.~{Walters} {\em et~al.},
{\em \physrep} {\bf 821}, 1
(2019).

\bibitem{Macquart:2020lln}
J.~P. {Macquart}, J.~X. {Prochaska}, M.~{McQuinn} {\em et~al.},
  {\em \nat} {\bf 581}, 391
(2020).

\bibitem{2001PhR...349..125B}
R.~{Barkana} and A.~{Loeb},
{\em \physrep} {\bf 349}, 125
(2001).

\bibitem{2009CSci...97..841C}
T.~R. {Choudhury},
{\em Current Science} {\bf 97}, 
841
(2009).

\bibitem{2013MNRAS.433..639P}
G.~{Paciga}, J.~G. {Albert}, K.~{Bandura} {\em et~al.},
{\em \mnras} {\bf 433}, 639
(2013).

\bibitem{2017ApJ...838...65P}
A.~H. {Patil}, S.~{Yatawatta}, L.~V.~E. {Koopmans} {\em et~al.},
  {\em \apj} {\bf 838}, 
  65
(2017).

\bibitem{2020MNRAS.493.1662M}
F.~G. {Mertens}, M.~{Mevius}, L.~V.~E. {Koopmans} {\em et~al.},
{\em \mnras} {\bf 493}, 1662
(2020).

\bibitem{2014PhRvD..89b3002D}
J.~S. {Dillon}, A.~{Liu}, C.~L. {Williams} {\em et~al.},
  {\em
  \prd} {\bf 89}, 
  023002
  (2014).

\bibitem{2015PhRvD..91l3011D}
J.~S. {Dillon}, A.~R. {Neben}, J.~N. {Hewitt} {\em et~al.},
  {\em \prd} {\bf 91}, 
  123011 
  (2015).

\bibitem{2016ApJ...833..102B}
A.~P. {Beardsley}, B.~J. {Hazelton}, I.~S. {Sullivan} {\em et~al.},
  {\em \apj} {\bf 833}, 
  102 
  (2016).

\bibitem{2016ApJ...833..213P}
S.~{Paul}, S.~K. {Sethi}, M.~F. {Morales} {\em et~al.},
  {\em \apj} {\bf 833}, 
  213 
  (2016).

\bibitem{2019ApJ...884....1B}
N.~{Barry}, M.~{Wilensky}, C.~M. {Trott} {\em et~al.},
  {\em \apj} {\bf 884}, 
  1
  (2019).

\bibitem{2019ApJ...887..141L}
W.~{Li}, J.~C. {Pober}, N.~{Barry} {\em et~al.},
  {\em \apj} {\bf 887}, 
  141
  (2019).

\bibitem{2020MNRAS.493.4711T}
C.~M. {Trott}, C.~H. {Jordan}, S.~{Midgley} {\em et~al.},
  {\em \mnras} {\bf 493}, 4711 
  (2020).

\bibitem{2019ApJ...883..133K}
M.~{Kolopanis}, D.~C. {Jacobs}, C.~{Cheng} {\em et~al.},
  {\em \apj} {\bf 883}, 
  133
  (2019).

\bibitem{2021arXiv210802263T}
{The HERA Collaboration},
  arXiv:2108.02263 
  (2021).

\bibitem{2018MNRAS.481.3821C}
T.~R. {Choudhury} and A.~{Paranjape},
  {\em \mnras} {\bf 481}, 3821 
  (2018).

\bibitem{2018Natur.555...67B}
J.~D. {Bowman}, A.~E.~E. {Rogers}, R.~A. {Monsalve}, T.~J. {Mozdzen} and
  N.~{Mahesh}, 
  {\em \nat} {\bf 555}, 67 
  (2018).

\bibitem{2018Natur.555...71B}
R.~{Barkana}, 
  {\em \nat} {\bf 555}, 71 
  (2018).

\bibitem{2018ApJ...858L..17F}
C.~{Feng} and G.~{Holder}, 
  {\em \apjl} {\bf 858}, 
  L17 
  (2018).

\bibitem{2019MNRAS.486.1763F}
A.~{Fialkov} and R.~{Barkana}, 
  {\em \mnras} {\bf 486}, 1763 
  (2019).

\bibitem{2020MNRAS.496.1445C}
A.~{Chatterjee}, P.~{Dayal}, T.~R. {Choudhury} and R.~{Schneider}, 
  {\em
  \mnras} {\bf 496}, 1445 
  (2020).

\bibitem{2018Natur.564E..32H}
R.~{Hills}, G.~{Kulkarni}, P.~D. {Meerburg} and E.~{Puchwein}, 
  {\em \nat} {\bf 564}, E32 
  (2018).

\bibitem{2021MNRAS.507.4684R}
J.~{Raste}, G.~{Kulkarni}, L.~C. {Keating} {\em et~al.},
  {\em \mnras}
  {\bf 507}, 4684 
  (2021).

\bibitem{planck2014-a12}
{Planck Collaboration},
  {\em \aap} {\bf 594}, 
  A10  (2016).

\bibitem{2005ApJ...622..759G}
K.~M. {G{\'o}rski}, E.~{Hivon}, A.~J. {Banday} {\em et~al.},
  {\em \apj} {\bf 622}, 759
  (2005).

\bibitem{SZ_1972}
R.~A. Sunyaev and Y.~B. Zel'dovich, 
  {\em Comments on Astrophysics and Space Physics} {\bf 4}, 
  173
  (1972).

\bibitem{Birkinshaw_1999}
M.~Birkinshaw, 
{\em \physrep} {\bf 310},
  97
  (1999).

\bibitem{Cen_1999}
R.~Cen and J.~P. Ostriker, 
{\em \apj} {\bf 514}, 
1
  (1999).

\bibitem{Tuominen_2021}
T.~Tuominen, J.~Nevalainen, E.~Tempel {\em et~al.}, 
  {\em \aap} {\bf 646}, 
  A156 
  (2021).

\bibitem{EBfullGrossmanpaper}
E.~S.~{Battistelli}, E. {Barbavara} and P.~{de Bernardis} {\em et~al.},
will appear in the {Proceedings of the MG16 Meeting on General Relativity}, online, 5 - 10 July 2021, edited by Remo Ruffini and Gregory Vereshchagin, World Scientific (2022).

\bibitem{Paiella_2019_olimpo}
A.~Paiella, E.~S. Battistelli, M.~G. Castellano {\em et~al.},
{\em Journal of Physics: Conference Series} {\bf 1182}, 
012005 
  (2019).

\bibitem{Masi2019}
S.~{Masi}, P.~{de Bernardis}, A.~{Paiella} {\em et~al.},
  {\em JCAP} {\bf 07}, 
  003
  (2019).


\bibitem{Petersen_2019}
M.~S. Petersen, R.~A. Gutermuth, E.~Nagel, G.~W. Wilson and J.~Lane,
  {\em MNRAS} {\bf 488}, 1462
  (2019).

\bibitem{Sokol}
A.~D. Sokol, R.~A. Gutermuth, R.~Pokhrel {\em et~al.},
  {\em MNRAS} {\bf 483}, 407 
  (2019).

\bibitem{Wall}
W.~F. Wall, I.~Puerari, R.~Tilanus {\em et~al.},
  {\em MNRAS} {\bf 459}, 1440 
  (2016).

\bibitem{Humphrey}
A.~Humphrey, M.~Zeballos, I.~Aretxaga {\em et~al.},
  {\em MNRAS} {\bf 418}, 74
(2011).

\bibitem{Planck2020VI}
{Planck Collaboration}, {\em A\&A} {\bf 641}, 
A6 
(2020).

\bibitem{Planck2014XX}
{Planck Collaboration}, {\em A\&A} {\bf 571}, 
A20 
(2014).

\bibitem{Planck2016XXIV}
{Planck Collaboration}, {\em A\&A} {\bf 594}, 
A24 
(2016).

\bibitem{Heymans2021}
C.~{Heymans} {\em et~al.}, {\em A\&A} {\bf 646}, 
A140
(2021).

\bibitem{Planck2016XXII}
{Planck Collaboration}, {\em A\&A} {\bf 594}, 
A22 
(2016).

\bibitem{Hurier2013}
G.~{Hurier} {\em et~al.}, {\em A\&A} {\bf 558}, 
A118 
(2013).

\bibitem{HTfullGrossmanpaper}
H.~{Tanimura}, M.~{Douspis} and N.~{Aghanim}, 
will appear in the {Proceedings of the MG16 Meeting on General Relativity}, online, 5 - 10 July 2021, edited by Remo Ruffini and Gregory Vereshchagin, World Scientific (2022).

\bibitem{Delabrouille2013}
J.~{Delabrouille} {\em et~al.}, {\em A\&A} {\bf 553}, 
A96 
(2013).

\bibitem{Maniyar2021}
A.~{Maniyar} {\em et~al.}, {\em A\&A} {\bf 645}, 
A40 
(2021).

\bibitem{planck2016-l01}
{Planck Collaboration},
  {\em \aap} {\bf 641}, 
  A1 
  (2020).

\bibitem{planck2013-pipaberration}
{Planck Collaboration},
  {\em \aap} {\bf 571}, 
  A27 
  (2014).

\bibitem{planck2020-LVI}
{Planck Collaboration},
  {\em \aap} {\bf 644}, 
  100  
  (2020).

\bibitem{RSfullGrossmanpaper}
R.~M. {Sullivan} and D.~{Scott},
will appear in the {Proceedings of the MG16 Meeting on General Relativity}, online, 5 - 10 July 2021, edited by Remo Ruffini and Gregory Vereshchagin, World Scientific (2022).

\bibitem{2021A&A...646A..75T}
T.~{Trombetti}, C.~{Burigana} and F.~{Chierici}, 
{\em \aap} {\bf 646}, 
A75
  (2021).

\bibitem{1997ApJ...482....6S}
U.~{Seljak}, 
{\em \apj} {\bf 482}, 6 
  (1997).

\bibitem{1997PhRvL..78.2058K}
M.~{Kamionkowski}, A.~{Kosowsky} and A.~{Stebbins},
{\em \prl} {\bf 78}, 2058 
  (1997).

\bibitem{1997PhRvL..78.2054S}
U.~{Seljak} and M.~{Zaldarriaga}, 
{\em \prl} {\bf 78}, 2054 
  (1997).

\bibitem{HuWhite97}
W. Hu and M. White,
{\em New Astronony} {\bf 2}, 323
  (1997).

\bibitem{2007NewAR..51..275D}
R. Durrer, 
{\em New Astronony Review} {\bf 51}, 275 
  (2007).

\bibitem{2014PhRvL.112s1303B}
C. Bonvin, R. Durrer and R. Maartens, 
{\em {\prl}} {\bf 112}, 191303 
  (2014).

\bibitem{2016A&A...594A..19P}
{Planck Collaboration} 
{\em {\aap}} {\bf 594}, A19
  (2016).

\bibitem{1981JETPL..33..532M}
V.~F. {Mukhanov} and G.~V. {Chibisov},
{\em Sov. Journal of Exp. and Theor. Phys. Lett.} {\bf 33}, 
532 
  (1981).

\bibitem{1982JETP...56..258M}
V.~F. {Mukhanov} and G.~V. {Chibisov},
{\em Sov. Journal of Exp. and Theor. Phys.} {\bf 56}, 
258 
  (1982).

\bibitem{1982PhLB..115..295H}
S.~W. {Hawking}, 
  {\em Physics Letters B} {\bf 115}, 295 
  (1982).

\bibitem{1982PhRvL..49.1110G}
A.~H. {Guth} and S.~Y. {Pi},
  {\em \prl} {\bf 49}, 1110 
  (1982).

\bibitem{1982PhLB..117..175S}
A.~A. {Starobinsky}, 
  {\em Physics Letters B}
  {\bf 117}, 175 
  (1982).

\bibitem{1983PhRvD..28..679B}
J.~M. {Bardeen}, P.~J. {Steinhardt} and M.~S. {Turner}, 
{\em \prd} {\bf 28}, 679 
  (1983).

\bibitem{1985JETPL..41..493M}
V.~F. {Mukhanov},
{\em Sov. Journal of Exp. and Theor. Phys. Lett.} {\bf 41}, 
493 
  (1985).

\bibitem{2007PhRvD..76d3005B}
N. Bevis, M. Hindmarsh, M. Kunz and J. Urrestilla, 
{\em {\prd}} {\bf 76}, 043005
  (2007).

\bibitem{2011PhLB..695...26G}
J. Garc{\'\i}a-Bellido, R. Durrer, E. Fenu, D.~G. Figueroa and M. Kunz, 
{\em Physics Letters B} {\bf 695}, 26
  (2011).

\bibitem{2014PhRvD..89h3512F}
E. Fenu, D.~G. Figueroa, R. Durrer, J. Garc{\'\i}a-Bellido and M. Kunz,
{\em {\prd}} {\bf 89}, 083512
  (2014).

%
\bibitem{1998PhRvD..58b3003Z}
M.~{Zaldarriaga} and U.~{Seljak}, 
{\em \prd} {\bf 58}, 
023003 
  (1998).

\bibitem{2016A&A...586A.133P}
{Planck Collaboration},
  {\em \aap} {\bf 586}, 
  A133
  (2016).

\bibitem{2018PhRvL.121v1301B}
{BICEP2 Collaboration}, {Keck Array Collaboration},
  {\em \prl} {\bf 121}, 
  221301
  (2018).

\bibitem{2020A&A...641A..10P}
{Planck Collaboration},
  {\em \aap} {\bf 641}, 
  A10
  (2020).

\bibitem{2015PhRvL.114j1301B}
{BICEP2/Keck Collaboration}, {Planck Collaboration},
{\em \prl} {\bf 114}, 
101301
  (2015).

\bibitem{2021A&A...647A.128T}
M.~{Tristram}, A.~J. {Banday}, K.~M. {G{\'o}rski} {\em et~al.},
  {\em \aap} {\bf 647}, 
  A128 
  (2021).

\bibitem{2020A&A...643A..42P}
{Planck Collaboration},
  {\em \aap} {\bf
  643}, 
  A42 
  (2020).

\bibitem{2008ApJ...676...10E}
H.~K. {Eriksen}, J.~B. {Jewell}, C.~{Dickinson} {\em et~al.}, 
  {\em \apj} {\bf 676}, 10 
  (2008).

\bibitem{2016A&A...596A.108P}
{Planck Collaboration},
  {\em \aap} {\bf
  596}, 
  A108 
  (2016).

\bibitem{Kosowsky:1994cy}
A.~Kosowsky, 
{\em Annals Phys.}
  {\bf 246}, 49 (1996).

\bibitem{Seljak:1996is}
U.~Seljak and M.~Zaldarriaga, 
{\em \apj} {\bf 469}, 437
  (1996).

\bibitem{Zaldarriaga:1996xe}
M.~Zaldarriaga and U.~Seljak,
{\em Phys. Rev. D} {\bf 55}, 1830  (1997).

\bibitem{Cooray:2002nm}
A.~Cooray, A.~Melchiorri and J.~Silk, 
{\em Phys. Lett. B} {\bf 554}, 1 (2003).

\bibitem{Bavarsad:2009hm}
E.~Bavarsad, M.~Haghighat, Z.~Rezaei {\em et~al.},
  {\em Phys. Rev. D} {\bf
  81}, 
  084035 (2010).

\bibitem{ModaresVamegh:2019nja}
S.~Modares~Vamegh, M.~Haghighat, S.~Mahmoudi and R.~Mohammadi,
{\em Phys. Rev. D} {\bf 100}, 
103024 (2019).

\bibitem{Haghighat:2019rht}
M.~Haghighat, S.~Mahmoudi, R.~Mohammadi, S.~Tizchang and S.~S. Xue,
  {\em Phys. Rev. D} {\bf 101}, 
  123016  (2020).

\bibitem{Tizchang:2018qpe}
S.~Tizchang, S.~Batebi, M.~Haghighat and R.~Mohammadi, 
in {\em Proceedings of The 39th Intern. Conf. on High Energy Physics (ICHEP2018)}, 
  p.142, Proceedings of Science (PoS) (2019).

\bibitem{STfullGrossmanpaper}
S.~{Tizchang}, R.~{Mohammadi} and S.~S. {Xue}, 
will appear in the {Proceedings of the MG16 Meeting on General Relativity}, online, 5 - 10 July 2021, edited by Remo Ruffini and Gregory Vereshchagin, World Scientific (2022).

\bibitem{Colladay:1996iz}
D.~Colladay and V.~A. Kostelecky,
{\em Phys. Rev. D} {\bf 55}, 6760 (1997).

\bibitem{Colladay:1998fq}
D.~Colladay and V.~A. Kostelecky,
{\em Phys. Rev. D} {\bf 58}, 
116002 (1998).

\bibitem{Kostelecky:1999mr}
V.~A. Kostelecky and C.~D. Lane,
{\em Phys. Rev. D} {\bf 60}, 
116010  (1999).


\bibitem{Kostelecky:2008ts}
V.~A. Kostelecky and N.~Russell,
in {\em Proceedings of the Fourth Meeting on CPT and Lorentz Symmetry}, p. 308, edited by V. Alan Kosteleck\'y, World Scientific (2008).

\bibitem{Planck:2015sxf}
{Planck Collaboration},
  {\em \aap} {\bf 594}, 
  A20 (2016).

\bibitem{Bertone:2004pz}
G.~Bertone, D.~Hooper and J.~Silk,
{\em \physrep} {\bf 405}, 279  (2005).

\bibitem{Clifton:2011jh}
T.~Clifton, P.~G. Ferreira, A.~Padilla and C.~Skordis,
{\em \physrep} {\bf 513}, 1  (2012).

\bibitem{Lovelock:1971yv}
D.~Lovelock,
{\em J. Math. Phys.} {\bf 12}, 498  (1971).

\bibitem{Milgrom1983a}
M.~Milgrom,
{\em \apj} {\bf 270}, 365  (1983).

\bibitem{Milgrom1983b}
M.~Milgrom,
{\em \apj} {\bf 270}, 371  (1983).

\bibitem{Milgrom1983c}
M.~Milgrom,
{\em \apj} {\bf 270}, 384  (1983).

\bibitem{BekensteinMilgrom1984}
J.~Bekenstein and M.~Milgrom,
  {\em \apj} {\bf 286}, 7  (1984).
    
\bibitem{Berezhiani:2015bqa}
L. Berezhiani and J. Khoury,
  {\em \prd} {\bf 92}, 103510 (2015).

\bibitem{FamaeyMcGaugh2011}
B.~Famaey and S.~McGaugh,
  {\em Living Reviews in Relativity} {\bf 15},
  10  (2012).

\bibitem{Sanders1997}
R.~H. Sanders,
{\em \apj} {\bf 480}, 492  (1997).

\bibitem{Bekenstein2004}
J.~D. Bekenstein,
 {\em Phys. Rev. D} {\bf 70}, 
 083509  (2004), [Erratum:  {\em Phys.
  Rev. D} {\bf 71}, 
  069901 (2005)].

\bibitem{SkordisEtAl2005}
C.~Skordis, D.~F. Mota, P.~G. Ferreira and C.~Boehm,
{\em Phys. Rev. Lett.} {\bf 96}, 
011301  (2006).

\bibitem{DodelsonLiguori2006}
S.~Dodelson and M.~Liguori,
  {\em Phys. Rev. Lett.} {\bf 97}, 
  231301  (2006).

\bibitem{SkordisZlosnik2019}
C.~Skordis and T.~Zlosnik,
{\em Phys. Rev. D} {\bf 100},
 104013  (2019).

\bibitem{ZlosnikFerreiraStarkman2006b}
T.~G. Zlosnik, P.~G. Ferreira and G.~D. Starkman,
{\em Phys. Rev. D} {\bf 74}, 
044037  (2006).


\bibitem{Skordis:2020eui}
C.~Skordis and T.~Zlosnik,
{\em Phys. Rev. Lett.} {\bf 127}, 161302 (2021).

\bibitem{Scherrer2004}
R.~J. Scherrer,
{\em Phys. Rev. Lett.} {\bf 93}, 
011301  (2004).

\bibitem{ArkaniHamedEtAl2003}
N.~Arkani-Hamed, H.-C. Cheng, M.~A. Luty and S.~Mukohyama, 
{\em JHEP} {\bf 05}, 
074  (2004).

\bibitem{SCH92}
P.~{Schneider}, J.~{Ehlers} and E.~E. {Falco}, {\em {Gravitational Lenses}}, Springer-Verlag Berlin Heidelberg (1992).

\bibitem{PIL10}
G.~L. {Pilbratt}, J.~R. {Riedinger}, T.~{Passvogel} {\em et~al.},
  {\em \aap} {\bf 518},
  L1 
  (2010).

\bibitem{CAR11}
J.~E. {Carlstrom}, P.~A.~R. {Ade}, K.~A. {Aird} {\em et~al.},
  {\em PASP} {\bf 123}, 
  568 
  (2011).

\bibitem{DUN20}
L.~{Dunne}, L.~{Bonavera}, J.~{Gonzalez-Nuevo}, S.~J. {Maddox} and
  C.~{Vlahakis},
  {\em \mnras} {\bf 498}, 4635 
  (2020).

\bibitem{GON17}
J.~{Gonz{\'a}lez-Nuevo}, A.~{Lapi}, L.~{Bonavera} {\em et~al.},
  {\em JCAP} {\bf 10}, 
  024
  (2017).

\bibitem{BON19}
L.~{Bonavera}, J.~{Gonz{\'a}lez-Nuevo}, S.~L. {Su{\'a}rez G{\'o}mez}  {\em et~al.},
{\em JCAP} {\bf 09}, 
021
  (2019).

\bibitem{ZHE05}
Z.~{Zheng}, A.~A. {Berlind}, D.~H. {Weinberg} {\em et~al.},
  {\em \apj} {\bf 633}, 791
  (2005).

\bibitem{BON20}
L.~{Bonavera}, J.~{Gonz{\'a}lez-Nuevo}, M.~M. {Cueli} {\em et~al.},
  {\em \aap} {\bf 639}, 
  A128 
  (2020).

\bibitem{CUE21}
M.~M. {Cueli}, L.~{Bonavera}, J.~{Gonz{\'a}lez-Nuevo} and A.~{Lapi},
  {\em \aap} {\bf 645}, 
  A126
  (2021).

\bibitem{1999MNRAS.308..119S}
R.~K. {Sheth} and G.~{Tormen},
{\em \mnras} {\bf 308}, 119  (1999).

\bibitem{2008ApJ...688..709T}
J.~{Tinker}, A.~V. {Kravtsov}, A.~{Klypin} {\em et~al.},
  {\em \apj} {\bf 688},
  709 
  (2008).

\bibitem{GON21}
J.~{Gonz{\'a}lez-Nuevo}, M.~M. {Cueli}, L.~{Bonavera}  {\em et~al.},
  {\em \aap} {\bf 646}, 
  A152 
  (2021).

\bibitem{BON21}
L.~{Bonavera}, M.~{M.~M. Cueli M.}, J.~{Gonz{\'a}lez-Nuevo J.} {\em et~al.},
  {\em \aap} {\bf 656}, 
  A99
  (2021).
  
\bibitem{LBfullGrossmanpaper}
L.~{Bonavera}, M.~M. {Cueli} and J.~{Gonz{\'a}lez-Nuevo},
will appear in the {Proceedings of the MG16 Meeting on General Relativity}, online, 5 - 10 July 2021, edited by Remo Ruffini and Gregory Vereshchagin, World Scientific (2022).

\end{thebibliography}

}

\end{document}